\documentclass[a4paper,11pt]{article}
\pdfoutput=1 

\usepackage{authblk}
\usepackage{jheppub} 

\usepackage[T1]{fontenc} 

\usepackage{amsmath}
\usepackage{amssymb}
\usepackage{amsthm}
\usepackage{physics}
\usepackage{bbm}
\usepackage{enumitem}
\usepackage{braket}
\usepackage{float}

\newtheorem*{lemma*}{Lemma}
\def\O{\mathcal{O}}

\theoremstyle{definition}

\theoremstyle{remark}

\title{\boldmath  The Reflected Entanglement Spectrum for Free Fermions}

\author[1,2]{Souvik Dutta,}
\author[1]{Thomas Faulkner,}
\author[1]{Simon Lin}

\affiliation[1]{Department of Physics, University of Illinois, Urbana-Champaign \\ 1110 W. Green St., Urbana IL 61801, USA.}
\affiliation[2]{Veritas Technologies, 2625 Augustine Drive, Santa Clara, CA 95054, USA}
\emailAdd{sdutta9@illinois.edu}
\emailAdd{tomf@illinois.edu}
\emailAdd{shanlin3@illinois.edu}

\abstract{We consider the reflected entropy and the associated entanglement spectrum for free fermions reduced to two intervals in $1+1$ dimensions.
Working directly in the continuum theory the reflected entropy can be extracted from the spectrum of a singular integral equation whose kernel
is determined by the known free fermion modular evolved correlation function. 
We find the spectrum numerically and analytically in certain limits. For intervals that almost touch 
the reflected entanglement spectrum approaches the spectrum of the thermal density matrix.
This suggests that the reflected entanglement spectrum is well suited to the task of extracting physical data of the theory 
directly
from the ground state wave function.
}
\begin{document}

\maketitle
\flushbottom

\section{Introduction and summary}

Entanglement entropy is now a central topic in the study of QFT \cite{Faulkner:2022mlp,Casini:2022rlv}. Entanglement entropy itself is UV divergent, so many of the derivations/proofs of important results involving entanglement entropy must play a delicate game of UV regularization \cite{Casini:2015woa}. We would like to study quantities insensitive to this regularization procedure. One approach is to study new quantities, different from entanglement entropy, that are well defined in the continuum limit. In favorable circumstances these UV finite quantities can be thought of as approaching a regularized version of entanglement entropy in certain limits.
A prominent example \cite{Casini:2015woa} is half the mutual information $I(A:B)/2$ of two spatial regions $A,B$ on a fixed Cauchy slice that are almost complement to each other, but leave a small finite corridor between the entangling surfaces.
The appropriate limit sends the size of the corridor to zero. 
Another quantity, the focus of this paper, is half the reflected entropy $S_R(A:B)/2$ which uses this same geometric setup for $A,B$ \cite{Dutta:2019gen}. 

An advantage of reflected entropy compared to the mutual information is that the reflected entropy is an actual von Neumann entropy of some density matrix - 
the reflected density matrix. Thus we can use the same regulator to study the reflected entanglement spectrum as a proxy for the regular entanglement spectrum. 
For a 2d (non-chiral) CFT the entanglement spectrum for $A$ a single interval of length $L$ was computed by Calabrese-Lafevre \cite{calabrese2008entanglement}. The result reads:
\begin{equation}
D(\lambda) =\delta(\lambda - \lambda_{\max})   +  \frac{\ln \lambda_{\max}^{-1}}{\lambda} f( \ln \lambda_{\max}^{-1} \ln \lambda_{\max}/\lambda) \qquad f(t) = I_1(2t^{1/2}) t^{-1/2} \theta(t)
\end{equation}
where $\lambda_{\max} =  (L/\epsilon)^{-c/6}$ and $I_1$ is the Bessel function.
This reproduces the expected Renyi entropies:
\begin{equation}
\exp(-S_n(A)/(n-1)) = \int d \lambda \lambda^{n} D(\lambda)
\end{equation}
computed in \cite{Holzhey:1994we,Calabrese:2004eu}.
There are several undesirable features to this formula. The spectrum depends on the UV cutoff $\epsilon$ and so will depend on the regularization procedure. The spectrum is continuous (aside from the single delta function at the edge of the continuum). It is also universal, only depending on the central charge of the CFT and none of the other CFT data such as the operator spectrum and OPE coefficients. In contrast the reflected entanglement spectrum is UV insensitive, discrete and depends on the operator spectrum and OPE coefficients of the CFT.

We now give a brief introduction to reflected entropy and the associated spectrum, see \cite{Dutta:2019gen} for further details.
Given a density matrix $\rho$ acting on a finite-dimensional Hilbert space $\mathcal{H}$,
one can form the \textit{canonical purification} $\ket{\sqrt{\rho}}$ by interpreting $\sqrt{\rho}$ as a state in the doubled Hilbert space
\begin{equation}
\label{canonical}
  \ket{\sqrt{\rho}} \in \text{End}(\mathcal{H}) = \mathcal{H}\otimes \mathcal{H}^*,
\end{equation}
where $\mathcal{H}^*$ is the dual of $\mathcal{H}$.
This doubled Hilbert space is equipped with the inner product $\Braket{\rho|\sigma} = \text{Tr}(\rho^\dag \sigma)$.
In the case where $\rho_{AB}\in \mathcal{H}_A\otimes \mathcal{H}_B$ is a bipartite density matrix,
the canonical purification lives in the space $(\mathcal{H}_A\otimes \mathcal{H}_{A^*}) \otimes (\mathcal{H}_B \otimes \mathcal{H}_{B^*}) \equiv \mathcal{H}_{AA^*BB^*}$ and one defines the \textit{reflected entropy} as the von Neumann entropy
\begin{equation}
  S_R(A:B) = S(AA^*)_{\ket{\sqrt{\rho_{AB}}}} = S_{\text{vN}}(\rho_{AA^*}),
\end{equation}
where $\rho_{AA^*} = \tr_{BB^*}\ket{\sqrt{\rho_{AB}}}\bra{\sqrt{\rho_{AB}}}$ is the reduced density matrix obtained by tracing over $\mathcal{H}_{BB^*}$. 
The reflected entanglement spectrum is simply the spectrum of $\rho_{AA^\star}$ and we claim this is discrete even in the continuum limit. 

Taking the continuum limit proceeds as follows. As we send $B \rightarrow A^c$, the complement
region to $A$, the reflected entropy reduces to twice the entanglement entropy which is now divergent.
This divergent behavior can be understood as the non-existence of a tensor factorization $\mathcal{H} \neq \mathcal{H}_A\otimes \mathcal{H}_{A^c}$ 
of the global Hilbert space without introducing a cutoff. This is an intrinsic property of type-III von Neumann algebras $\mathcal{A}_A$ that govern
the local bounded operators associated to region $A$ \cite{Witten:2018zxz}. 
Keeping a finite gap between $A$ and $B$ however allows reflected entropy to be used as a regulated version of entanglement entropy.
In particular for two disjoint regions $A\cup B$, the ``split property'' \cite{cmp/1103859773,cmp/1104115703} guarantees the existence of at least one type-I factor $\mathcal{N}$ splitting of the local algebras
\begin{equation}
  \mathcal{A}_A \subset \mathcal{N} \subset \mathcal{A}_B'.
\end{equation}
The canonical purification introduced in \eqref{canonical}, in the algebraic language, corresponds to the state induced on a canonical type-I splitting factor, which can be written algebraically as \cite{Doplicher1984}
\begin{equation}
  \mathcal{N} = \mathcal{A}_A \vee J_{AB} \mathcal{A}_A  J_{AB},
\end{equation}
where $J_{AB}$ is the anti-unitary Tomita-Takesaki modular conjugation operator associated to $\mathcal{A}_{AB}$ and the vacuum state.
In this language the reflected entropy is defined as the von Neumann entropy of this type-I factor.
In particular, the density matrix of $\mathcal{N}$ is trace-class, with a well-defined and discrete spectrum \cite{reed1981functional}, allowing one to make sense of the entropy and spectrum directly in the continuum.

It was shown in \cite{Dutta:2019gen} that in the AdS/CFT setting, the reflected entropy is dual to the area of entanglement wedge cross-section \cite{Takayanagi:2017knl}.
Since the entanglement wedge cross-section is typically a deep bulk probe of the emergent geometry, it is interesting to study the reflected entropy in a more general class of QFTs.
Indeed, there has been many previous works computing the reflected entropy on different quantum systems, such as free fermions \cite{Bueno:2020vnx},  free scalars \cite{Bueno:2020fle}, CFT in arbitrary dimensions \cite{Camargo:2021aiq}, 3D Chern-Simons \cite{Berthiere:2020ihq}, holographic tensor networks \cite{Akers:2021pvd} and JT gravity with EOW branes \cite{Akers:2022max}.

In this note we will focus on free fermions in $1+1$ dimensions. 
Reflected entropy in free fermion systems was also already studied in \cite{Bueno:2020vnx,Longo:2019pjj} (see also upcoming work \cite{upcoming}). In this paper we will make a few new observations. 
In particular compared to \cite{Bueno:2020vnx} we will work directly in the continuum theory bypassing the need of discretization. Some of our results will be numerical although we will also make some new analytic predictions in various limits. 

\begin{figure}[h]
  \centering
  \includegraphics[width=.8\textwidth]{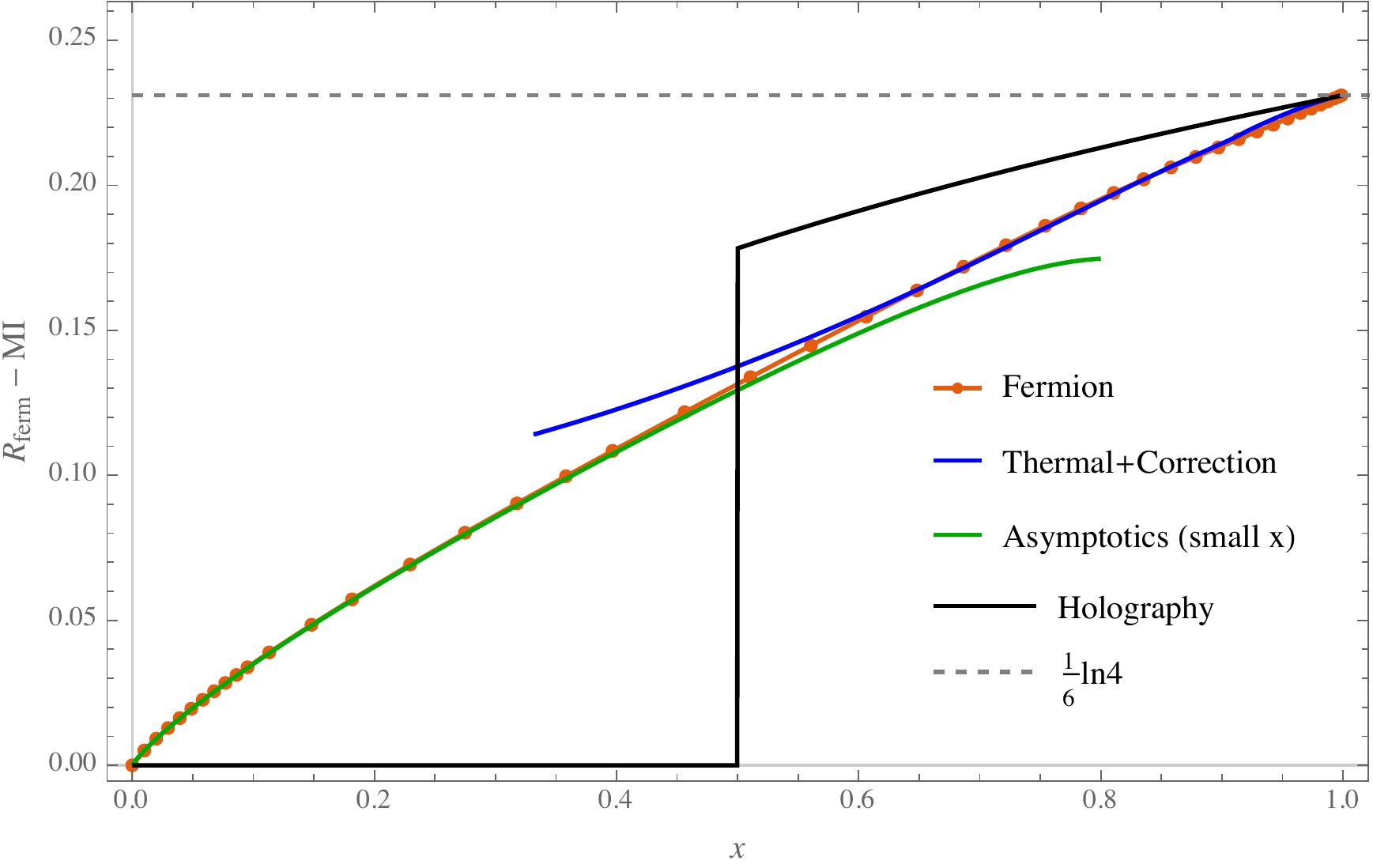}
  \caption{Numerical result of free fermion reflected entropy. We plot here the markov gap \cite{Hayden:2021gno} $S_R-\text{MI}(A:B)$ (where $\text{MI}$ is the mutual information) of the free Fermion.  In contrast to the reflected entropy which is divergent as $x\to 1$, the markov gap is finite and approaches $1/6 \ln 4$ as $x\to 1$, as implied from the general behavior of 2D CFT \cite{Dutta:2019gen}.
   This asymptotic value is shown as the gray dashed line in the figure. 
    The orange dots are obtained from numerical approximations of the correlation kernel \eqref{eq:C_mn}.
    The asymptotic formula at $x\to 0$ is obtained from perturbation theory around small $x$ \eqref{eq:asym_smallx}.
    The asymptotic formula at $x\to 1$ is the entropy calculated from thermal distribution \eqref{eq:thermal_Z} plus next order corrections \eqref{eq:lambdashift}. 
    We have also included the reflected entropy of a holographic CFT (divided by $2c$ for comparison), which undergoes a phase transition in which its value jumps from $O(c^0)$ to $O(c)$ at $x=1/2$.
    }
  \label{fig:Rferm}
\end{figure}

Consider two intervals $A$ and $B$ on some equal time slice in vacuum and define the cross-ratio of the end points of these intervals as $x$.
We will give analytic predictions for $S_R$ in the limit where the $A$ and $B$ intervals are far separated $ x \rightarrow 0$ or nearly touching $x \rightarrow 1$. In the former case the reflected entropy behaves as:
\begin{equation}
S_R = \alpha (-x \ln x) + \beta + \ldots 
\end{equation} 
where we give a simple integral expression \eqref{eq:alpha_int}  for $\alpha$ and a numerical prediction \eqref{eq:beta_int} for $\beta$.
This is in agreement with our numerics as well as the numerics in \cite{Bueno:2020vnx} and qualitative agreement with \cite{Camargo:2021aiq}.
In the later case we show that to the leading order
\begin{equation}
S_R =  \frac{1}{6}\left( - \ln (1-x) + \ln 4 \right) + \ldots
\end{equation} 
in agreement with the universal behavior of 2D CFTs \cite{Dutta:2019gen}.
In addition we give the next order correction to $S_R$, agreeing well with our numerics.
We summarize the analytical predictions and numerics in Figure~\ref{fig:Rferm}.

\begin{figure}[h]
  \centering
  \includegraphics[width=.7\textwidth]{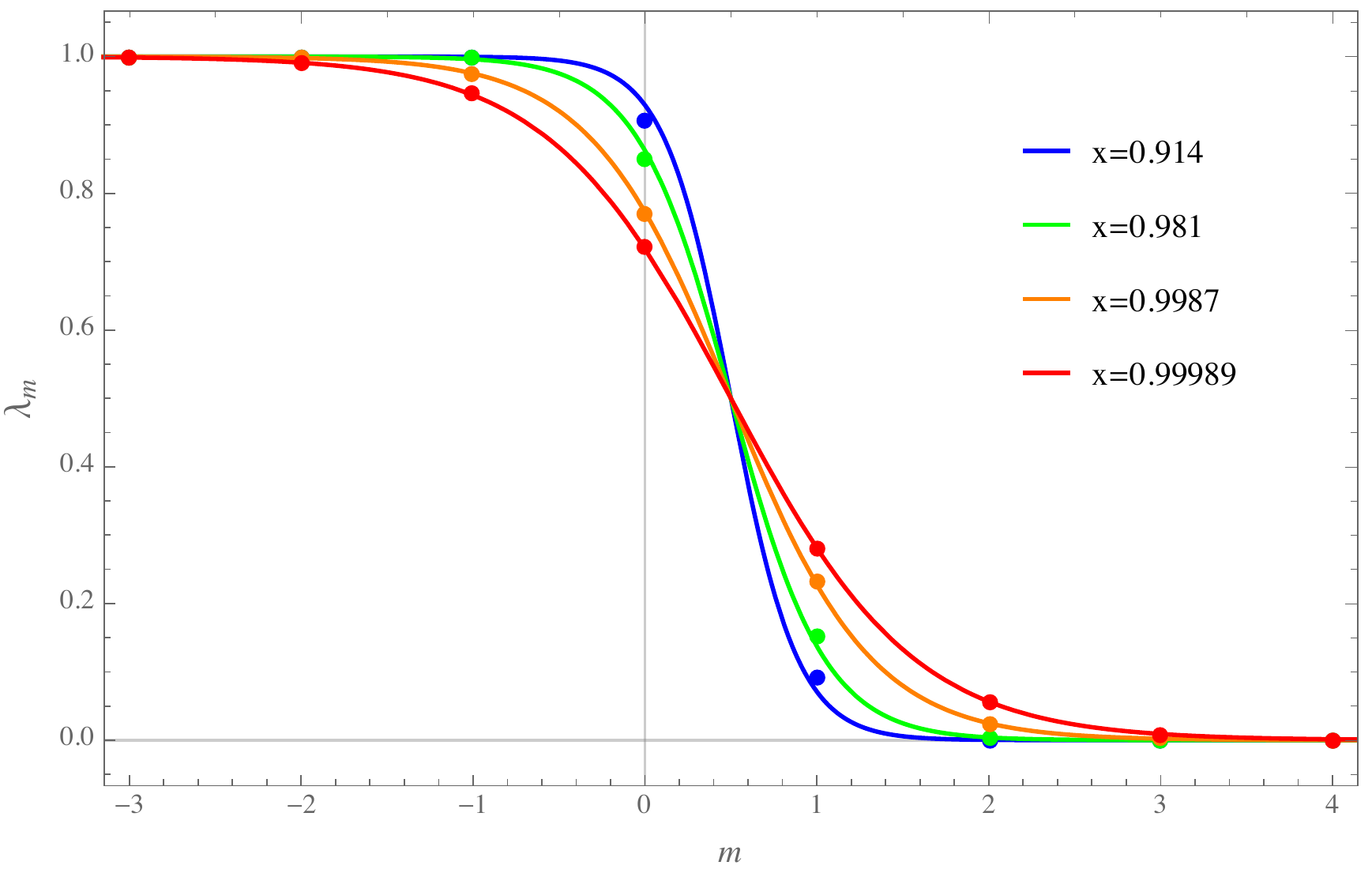}
  \caption{Eigenvalues $\lambda_m$ of the correlation kernel obtained by numerical method (dots) versus the spectrum \eqref{eq:lambdaleading} of a thermal partition function with appropriate temperature (solid lines) for various $x\to 1$. Only the first few eigenvalues are shown. $\lambda_m$ are symmetric across $m=1/2$ line.
  Although not shown in this figure, the discrepancy between analytics and numerics is well resolved by including the second order corrections, see Figure~\ref{fig:lambdashift}.}
  \label{fig:eigenvalue}
\end{figure}

We also derive analytically the entanglement spectrum of $\rho_{AA^\star}$ as $x \rightarrow 1$.
It takes the simple form of that of the spectrum of the thermal density matrix for a free chiral fermion on the circle in the NS sector with inverse temperature to circle length ratio  $\beta/L \propto  -1/\ln(1-x)$. 
By studying perturbative corrections to this later spectrum we give evidence that the reflected density matrix approaches rapidly the thermal density matrix.
This agrees well with the eigenvalues obtained from numerical method, see Figure~\ref{fig:eigenvalue}.

In this paper we do not make use of the replica trick, as in previous work \cite{Bueno:2020vnx}, but we use the correlation matrix technique \cite{Casini:2009sr} that is based on the fact that a many-body Gaussian state is entirely determined by it's two point correlation function.  And so the many-body entropy can also be extracted from this
correlation function.
It is sufficient to know certain modular flowed two point functions in order to construct the relevant correlation matrix that computes the reflected entropy. These modular flow correlators were computed in the continuum in \cite{Casini:2009vk, Longo:2009mn, Hollands:2019hje}. 

Our analytic computations proceed as follows. 
We setup a systematic expansion for the eigenfunctions of the correlation matrix in the limit $x \rightarrow 1$ via a certain matching procedure, between the endpoints
of $A$. This is similar to a  QM scattering problem. 
In this limit one starts out with a continuum of  Rindler eigenfunctions near each $A$ endpoint and the discrete spectrum arises from a matching condition in-between.
The discreteness of the spectrum is important for the finite-ness of the reflected entropy.

The plan of this paper is as follows. We setup the singular integral equations in Section~\ref{sec:setup}. In Section~\ref{sec:zero} and Section~\ref{sec:one} we discuss the $x \rightarrow 0$ limit 
and $x \rightarrow 1$ limits respectively. In the later limit we make a more careful study of the spectrum, computing sub-leading corrections to the thermal spectrum in Section~\ref{sec:corrections}. We also introduce a $s$-modular flowed version of reflected entropy, that we call deflected entropy in Section~\ref{sec:deflected} - this is a natural one parameter generalization of reflected entropy. We compute this quantity for free fermions and also in AdS$_3$/CFT$_2$ using the methods of \cite{Faulkner:2013yia} - the dual roughly corresponds to a reflected geodesic that tracks the entanglement wedge cross section but picks up a boosts of rapidity $s$ at the Ryu-Takayanagi surface. 
 In Section~\ref{sec:discussion} we discuss our results, and make some conjectures about more general CFTs. 
Numerical results will be presented throughout the paper but are summarized in Figure~\ref{fig:Rferm}.

\section{Setup}
\label{sec:setup}

\begin{figure}[h]
  \centering
  \includegraphics[width=0.5\textwidth]{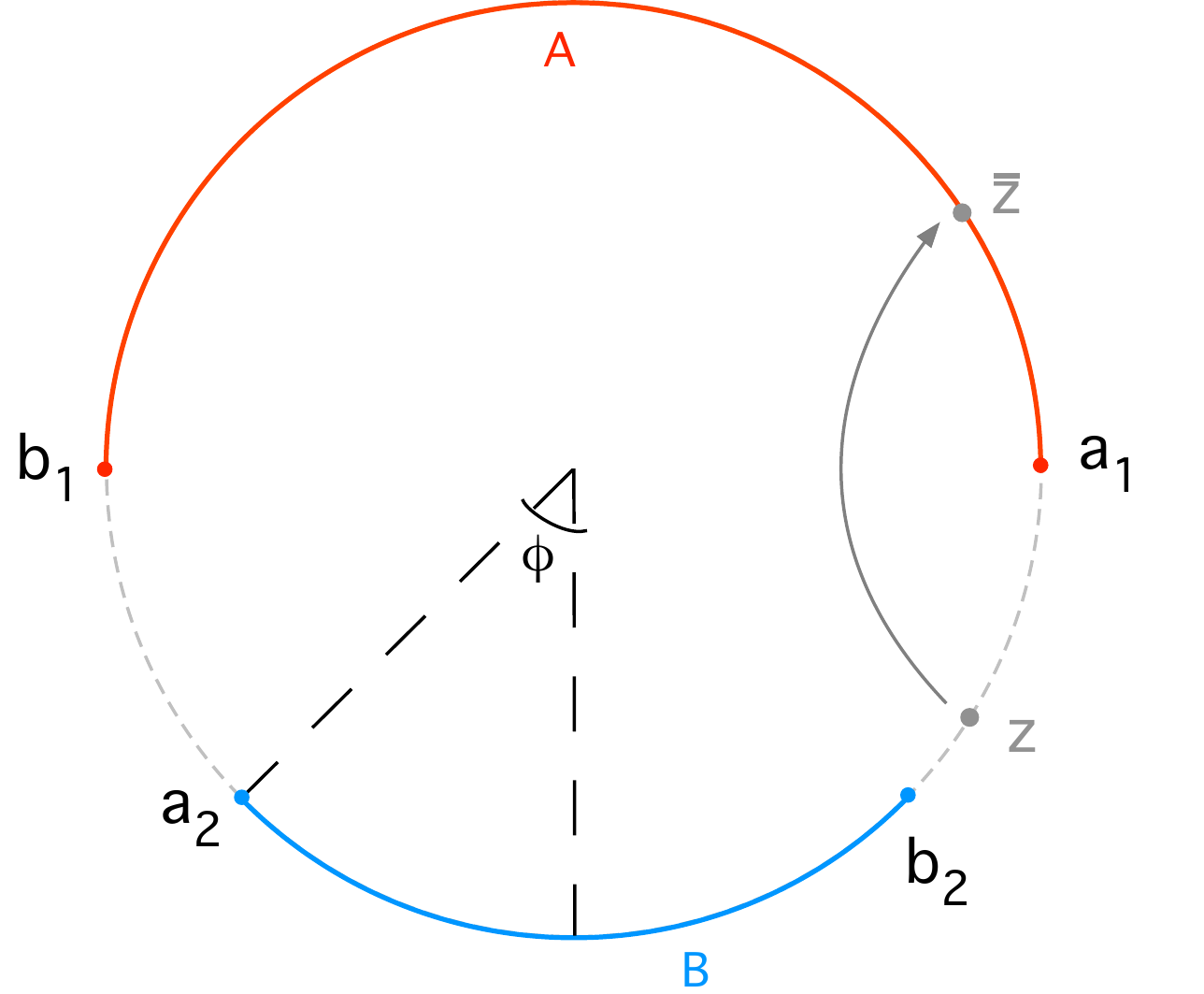}
  \caption{
    Our setup for calculating the reflected entropy for a free fermion on a circle.
    Global conformal symmetry allows us to fix three of the four the interval endpoints $A = [a_1,b_1] = [0,\pi]$ and $B=[a_2,b_2] = [3\pi/2-\phi,3\pi/2+\phi]$. The angle $\phi$ is related to the conformal cross-ratio by $x = 2\sin\phi/(1+\sin\phi)$. The modular conjugation for region $A$ takes $z\to \bar{z}$, or equivilently $\theta\to -\theta$ after circular identification.}
  \label{fig:setup}
\end{figure}

We would like to compute the reflected entropy for 2d free chiral fermions $\psi$ and two intervals $A,B$ on a circle.
As discussed in \cite{Bueno:2020vnx}, the reflected density matrix is still Gaussian (that is the Fermion fields satisfy Wick's theorem) so that the reduced density matrix is completely determined by the two point correlation function of fundamental fermions.
We can then infer the entropy by solving the singular integral equation of this correlation kernel \cite{Casini:2009sr}.
The entire system consists of Fermions in $AB$ and in the doubled system $(AB)^\star$.
This later system is governed by Fermionic operators $\widehat{\psi}(x) =  i \tilde{J}_{AB} \psi(x)^\dagger \tilde{J}_{AB}^\dagger$ for $x\in AB$.
Since Fermions anti-commute, we need a generalization of the Tomita-Takesaki theory to a graded algebra \cite{Longo:2017mbg,Longo:2019pjj,Bueno:2020vnx}.
Denoting $\Gamma = (-1)^F$ where $F$ is the Fermion number operator,
we can form the Klein operator $Z = \frac{1-i\Gamma}{1-i}$, which is a unitary operator that acts as $1$ on even states and $i$ on odd states.
The modular conjugation operator $\tilde{J}_{AB}$ for the graded system is then related to the vacuum modular conjugation of Tomita-Takesaki theory (for the full fermionic algebra of
operators in $AB$) $J_{AB}$ by $\tilde{J}_{AB} = Z J_{AB} = J_{AB} Z^\dagger$.
The conjugated modes $\widehat{\psi}(x)$ then satisfy the canonical anti-commutation relations amongst themselves as well as with the original $AB$ system. We are interested in the entanglement entropy of $AA^\star$ so we may restrict to correlation functions of $\psi|_A$ and $\widehat{\psi}|_A$. 

In order to understand the smooth $B \rightarrow \emptyset$ limit we will choose to geometrize the  $A^\star$ reflected fermions by using instead the following description:
\begin{equation}
  \widetilde{\psi}(x) \equiv \tilde{J}_{AB} \tilde{J}_A \psi(x) \tilde{J}_A^\dagger \tilde{J}_{AB}^\dagger  = J_{AB} J_A \psi(x) J_A J_{AB} \,, \qquad x \in A^c
\end{equation}
which is a fermion that now lives on $A^c$ where $A^c$ is the complementary region to $A$. The $\tilde{J}_A$ operator acts as CPT conjugation with a reflection across the $A$ entangling surface,
thus these still represent the same modes as $\widehat{\psi}|_A$. The system $AA^\star$ has now been mapped to the entire circle: $S^1 = AA^c$.
If we additionally define  $\widetilde{\psi}(x)|_{A} \equiv \psi(x)$ then $\widetilde{\psi}$ is the new Fermion that lives on this circle:
it satisfies the standard canonical anti-commutation relations.
We thus, simply need to work out the state of the $\widetilde{\psi}$ fermion on this circle. As discussed above this is determined by the correlation function of $\widetilde{\psi}$. Note that we expect
this to be an appropriately smooth state since near $\partial A$ one can check that
$J_{AB}$ acts geometrically like the Rindler reflection or $J_A$, so the two modular operators cancel out. 

Notice  that the $B \rightarrow \emptyset$ now reproduces the Fermion on a circle with the vacuum state.
This is as expected and will give zero reflected entropy. 
The free fermion correlator on the cylinder in the NS vacuum $\left| \Omega \right>$ is the following distribution:
\begin{equation}
\left< \Omega \right| \psi^\dag(x)\psi(y) \left| \Omega \right>  =  
 \mathcal{P} \frac{z^{1/2}w^{1/2}}{\pi (z-w)} 
 + \frac{1}{2}\delta(x-y) \, \qquad z= e^{ix}, \quad w = e^{iy}
\end{equation}
where $ 0 < x,y \leq 2\pi$ on the cylinder. We have the canonical anti-commutation relations $\{ \psi^\dag(x), \psi(y) \} = \delta(x-y)$ (with all other anti-commutators vanishing).  
Consider a basis of normalized functions on the circle that is anti-periodic: $e^{ i (m+1/2)x } (2\pi)^{-1/2}$. 
We can consider the modes:
\begin{equation}
\psi_m =\frac{1}{\sqrt{2\pi}} \oint \frac{dz}{ i z } z^{m+1/2}  \psi(x)\, \qquad m \in \mathbb{Z}
\end{equation}
which gives $\{ \psi_n^\dagger , \psi_m \} = \delta_{m,n}$ from the canonical anti-commutation relations. 

The Fourier modes of this correlator ($m,n\in \mathbb{Z}$) can then be calculated:
\begin{equation}
  \label{eq:theta_m}
  C_{mn} = \oint \frac{dz}{z i \sqrt{2\pi}} z^{m+1/2} \oint \frac{dw}{w i \sqrt{2\pi}} w^{-(n+1/2)}\left< \Omega \right| \psi^\dag(x)\psi(y) \left| \Omega \right>\\
  = \Theta_m\delta_{mn}
\end{equation}
where $\Theta_m = 1$ if $m\ge0$ and $0$ otherwise.
Thus we find that the correlation function/matrix becomes a projector with  eigenvalues $0$ and $1$.

Let us now compute the correlation function when $B\neq\emptyset$:
\begin{equation}
  \label{eq:correlator}
{\rm Tr}_{AA^c} \left( \rho_{ AA^c} \widetilde{\psi}^\dagger (x) \widetilde{\psi}(y)  \right)
  = \begin{cases} \left< \Omega \right| \psi^\dagger (x) \psi(y) \left| \Omega \right> & x,y \in A \quad {\rm or}  \quad x, y \in A^c  \\
    \left( \frac{\partial y}{\partial y_J} \right)^{1/2}  \left< \Omega \right| \psi^\dagger (x) \Delta_{AB}^{1/2} \psi(y_J) \left| \Omega \right> & x \in A\,,\quad y \in A^c \\
    - \left( \frac{\partial x}{\partial x_J} \right)^{1/2}  \left< \Omega \right| \psi (y) \Delta_{AB}^{1/2} \psi^\dagger(x_J) \left|\Omega \right> & y \in A\,,\quad x \in A^c \\
  \end{cases}
\end{equation}
where $x_J$ is given by the geometric action of $\tilde{J}_A$ on $x$: if $x \in A$ then $x_J \in A^c$. 
We have used Tomita-Takesaki theory \cite{Takesaki:1970aki} to replace the modular conjugation operator $J_{AB}$ with the modular operator $\Delta_{AB}$
for the vacuum state.
The phase of the square root can be fixed by demanding we reproduce the vacuum correlator as $B \rightarrow \emptyset$.
This correlation function along with the Gaussian nature of $\rho_{AA^c}$ will allow us to compute its spectrum.
The spectrum of $\rho_{AA^c}$ is the same as the spectrum of the reflected density matrix $\rho_{AA^\star}$ or equivilently the density matrix on the canonical type-I factor $\mathcal{N}$. 

The correlation functions can be computed from the formula for modular flow for fermions on the plane \cite{Casini:2009vk, Longo:2009mn, Hollands:2019hje}:
\begin{align}
  \left< \psi (z) \Delta_{AB}^{is} \psi^\dagger(w) \right> &=  \left< \psi^\dagger (z) \Delta_{AB}^{is} \psi(w) \right> = \frac{e^{-\pi s}}{2\pi i (z - w)}  \left( 1 - 
                                            \frac{1 - e^{-2\pi s}}{ Q - e^{-2\pi s}} \right)\, \\ \qquad Q &= \frac{(z-a_1)(z-a_2)(w-b_1)(w-b_2)}{ (z-b_1)(z-b_2)(w-a_1)(w-a_2)}
\end{align}
where $A = [a_1, b_1]$ and $B = [a_2,b_2]$ and we may consider the boundaries of these intervals to be complex coordinates in the plane. 

For Fermions on the circle we are free to fix $a_1 = 1, b_1 =-1$ and $a_2 = -i e^{-i \phi}, b_2 = -i e^{i \phi}$
and set $z = e^{i x}$ and $w = e^{i y}$. Here we may consider $x$ as a holomorphic coordinate on the cylinder with $0 \leq {\rm Re} \, x  \leq 2\pi$ (not to be confused with the cross-ratio that we will introduce later)
and in these coordinates $A = [0,\pi]$ and $B = [ 3\pi/2-\phi, 3\pi/2+\phi]$ (see figure~\ref{fig:setup}). We use:
\begin{equation}
  \left< \psi^\dagger (x) \Delta_{AB}^{is} \psi(y) \right>_\Omega \equiv  \left< \psi^\dagger (x) \Delta_{AB}^{is} \ \psi(y) \right>_{\rm cyl}  = \left( \frac{ \partial z}{\partial x} \right)^{1/2}  \left( \frac{ \partial w}{\partial y} \right)^{1/2} \left< \psi^\dagger (x) \Delta_{AB}^{is} \ \psi(y) \right>
\end{equation}
and $x_J = - x$.
If we subtract the vacuum correlator we find:
\begin{align}
  \left< \widetilde{\psi}^\dagger (x) \widetilde{\psi}(y) \right>_{\rho_{AA^c}} &- \left< \psi^\dagger (x) \psi(y) \right>_\Omega
  \\ &=  \frac{1}{4 \pi i \sin(\frac{x-y}{2})} \frac{ \theta_A(x) \theta_{A^c}(y)  \sin(x) \sin(y) \sin(\phi) }{ ( \sin^2(\frac{x-y}{2}) - \sin\phi  \sin^2(\frac{x+y}{2})  + \cos\phi \sin(\frac{x-y}{2}) \cos( \frac{x + y}{2} ) )} \nonumber \\
                                                                 &   \qquad \qquad \qquad \qquad \quad - (x \leftrightarrow y) \nonumber
\end{align}
where $ \theta_A(x) ( \theta_{A^c}(x)) $ is a unit step for $x\in A (A^c)$.
Define the Fourier transform of the correlation kernel
\begin{equation}
  \Delta C_{mn} = \int_0^{2\pi}  \frac{d x}{\sqrt{2\pi}} e^{i x(m+1/2) } \int_0^{2\pi}  \frac{d y}{\sqrt{2\pi}} e^{-i y(n+1/2) }\left( \left< \widetilde{\psi}^\dagger (x) \widetilde{\psi}(y) \right>_{\rho_{AA^\star}} - \left< \psi^\dagger (x) \psi(y) \right>_\Omega
  \right)
\end{equation}
So that we have
\begin{equation}
  \label{eq:C_mn}
  \Delta C_{mn} 
  = \int_0^{\pi}  dx  \int_\pi^{2\pi}  dy
  \frac{ 1}{4 \pi^2  \sin(\frac{x-y}{2})} \frac{ \sin( x(m+1/2) - y(n+1/2))  \sin(x) \sin(y) \sin(\phi) }{ ( \sin^2(\frac{x-y}{2}) - \sin\phi  \sin^2(\frac{x+y}{2})  + \cos\phi \sin(\frac{x-y}{2}) \cos( \frac{x + y}{2} ) )}
\end{equation}
from which it is clear that $\Delta C_{mn}  = \Delta C_{nm}$ (from sending $x \leftrightarrow 2\pi -y$).
Furthermore $x \rightarrow \pi - x$ and $y \rightarrow 3\pi - y$ gives:
\begin{equation}
  \Delta C_{mn} = (-1)^{m-n} \Delta C_{mn}
\end{equation}
which implies that $\Delta C_{mn}$ vanishes unless $m-n$ is even.  Also it is clear that $\Delta C_{m,n} = -\Delta C_{-1-m,-1-n} $.

The actual correlation matrix is:
\begin{equation}
  C_{mn} = \Delta C_{mn} + \delta_{mn} \Theta_m
\end{equation}
with $\Theta_m$ defined in \eqref{eq:theta_m}.
The second term comes from the vacuum correlator (this is the projector which gave
zero entanglement when $ \Delta C_{mn}  = 0$.)
The reflected entropy can be computed via \cite{Casini:2009sr}
\begin{align}
  S_R = -\tr (C_{mn}\ln C_{mn}) -\tr((1-C_{mn})\ln(1-C_{mn})) 
\end{align}

\begin{figure}[h]
  \centering
  \includegraphics[width=.6\textwidth]{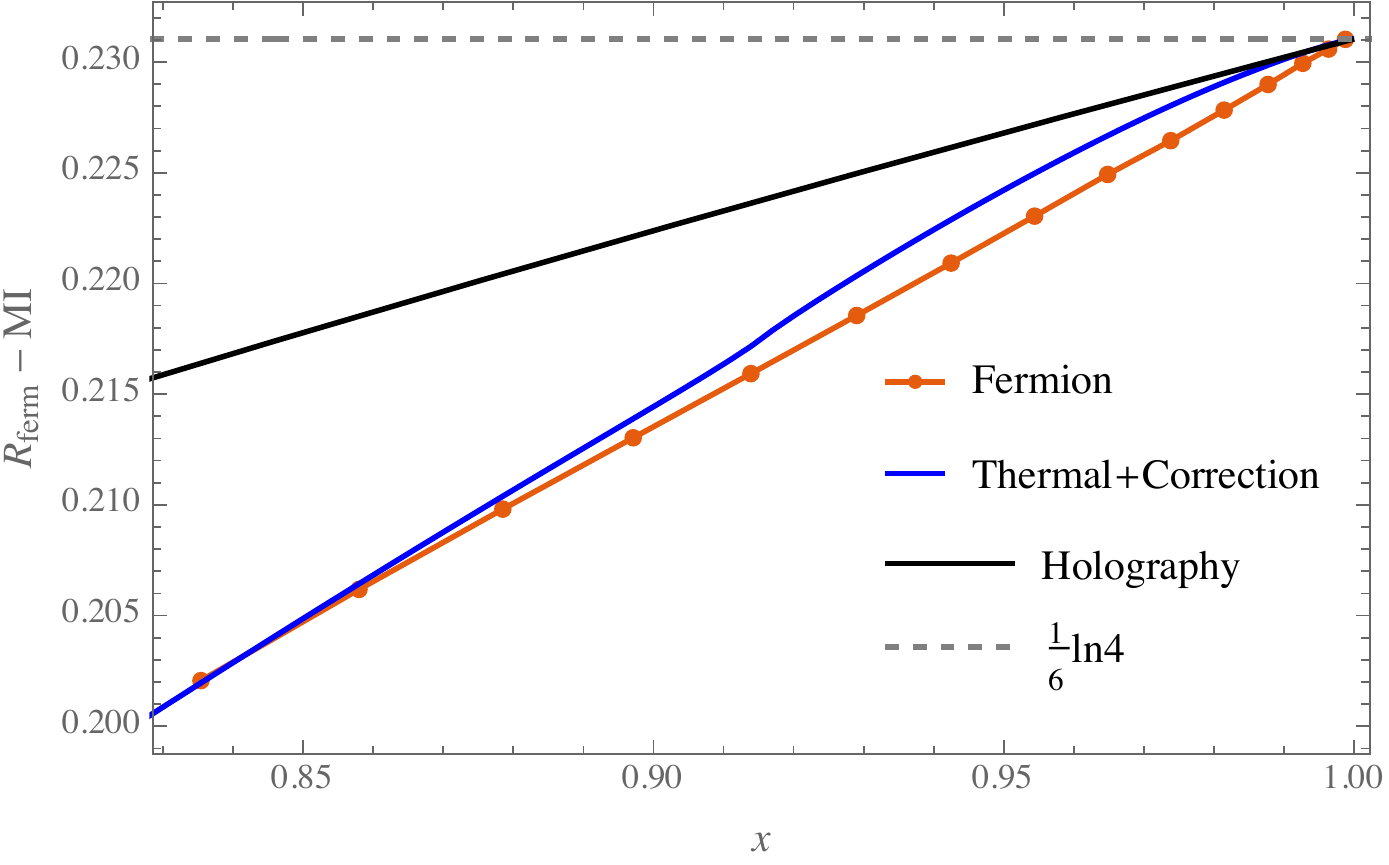}
  \caption{The Markov gap $S_R-\text{MI}$ of the analytics versus results obtained from numerics at large conformal cross-ratio $x$.
    The analytical prediction is based on the leading order thermal prediction \eqref{eq:lambdaleading}, plus the corrections \eqref{eq:lambdashift} obtained in this subsection.
    This figure is the magnified version of Figure~\ref{fig:Rferm} at $x\to 1$.}
  \label{fig:Rferm_large_x}
\end{figure}

After truncation of the modes we can numerically compute $\Delta C_{mn}$ and use this to extract the reflected entropy (Figure~\ref{fig:Rferm},\ref{fig:Rferm_small_x},\ref{fig:Rferm_large_x},\ref{fig:s_plots}) and entanglement spectrum (Figure~\ref{fig:eigenvalue},\ref{fig:lambdashift}).
The relevant cross ratio is:
\begin{equation}
  x = \frac{ 2 \sin\phi}{1 + \sin\phi}
\end{equation}
We truncate the matrix at some finite value $|m|,|n|\le L$ so that the resulted entropy converge to prescribed tolerance.
\footnote{The appropriate value of $L$ where we see convergence is highly dependent on the cross-ratio $x$ as well as desired precision. In general we need higher $L$ when $x\to 1$. For the plots in this note the value of $L$ ranges from $20$ to $500$.}

Starting from the next section we will study the reflected entropy under various limits.
Clearly the limit $\phi \rightarrow 0$ ($x\to 0$) is most naturally studied on the cylinder; whereas the limit $\phi \rightarrow \pi/2$ ($x\to 1$) is most appropriately studied using Rindler space, since this is the limit where we recover the usual (divergent) entanglement entropy.

\section{The $x\to 0$ limit}

\label{sec:zero}

Let us first consider the case when $\phi\to0$.
We will apply degenerate perturbation theory to the correlation matrix, which allows us to extract the $\phi$ and $\phi\ln\phi$ terms of the reflected entropy.
Taylor expand \eqref{eq:C_mn} to the first order in $\phi$ we have:
\begin{align}
  \Delta C_{mn}
  & \approx \phi {\rm Re}  \int_0^\pi \frac{d x}{\pi} \frac{e^{i x m } \sin(x)}{1+i e^{-i x}}
    \int_\pi^{2\pi} \frac{d y}{\pi} \frac{e^{-i y n } \sin(y)}{1+i e^{i y}} \frac{ 1}{2\sin^2(\frac{x-y}{2})}
  \\
  & = (-1)^n  \phi {\rm Re}  \int_0^\pi \frac{dx}{\pi} \int_0^\pi \frac{dy}{\pi}  \frac{f_m(x) f_n(y)^\star}{2\cos^2(\frac{x-y}{2})}
    \,, \quad f_m(x) = \frac{e^{i x m} \sin(x)}{ 1 + i e^{- ix}}
\end{align}

\begin{figure}[h]
  \centering
  \includegraphics[width=.6\textwidth]{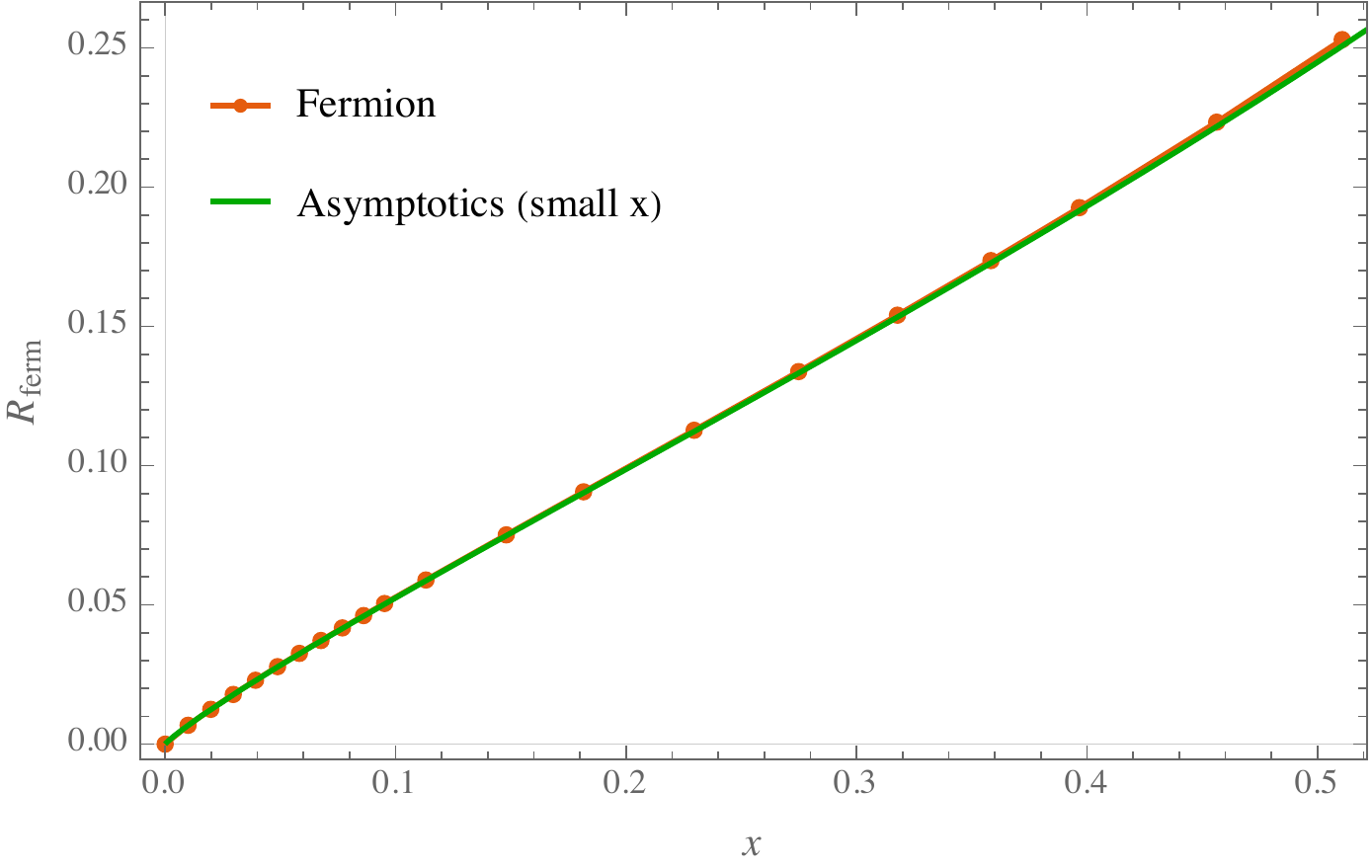}
  \caption{The reflected entropy for the analytical prediction \eqref{eq:asym_smallx} versus the numerics.    
    Numerical fitting of the data points with $x<0.1$ to $S_R = -\alpha\phi\ln\phi+\beta\phi$ gives $\{\alpha,\beta\} = \{0.149,0.563\}$, agreeing well with the expression we obtained in \eqref{eq:alpha_int} and \eqref{eq:beta_int}.  }
  \label{fig:Rferm_small_x}
\end{figure}

Since the eigenvalue is degenerate at zeroth order, we need to diagonalize the two matrices on these degenerate subspaces:
\begin{equation}
  A = \lim_{\phi \rightarrow 0} \frac{1}{\phi} \Theta \Delta C \Theta \, \qquad B = \lim_{\phi \rightarrow 0} \frac{1}{\phi} (1-\Theta) \Delta C (1- \Theta)
\end{equation}
where we defined the projector $\Theta_{mn} = \delta_{mn} \Theta_m$. Define the eigenvalues of these matrices
as $-a_k , b_k$ respectively such that $a_k, b_k  \geq 0$. 
Then the eigenvalues of $C_{mn}$ are approximately $(1- \phi a_k)$ and $ \phi  b_k $ to first order in $\phi$. The leading correction to the reflected entropy is:
\begin{align}
  \label{eq:asym_smallx}
  S_R &\approx \sum_k \left(  \phi a_k  (1- \ln\left( \phi a_k \right))  + \phi b_k  (1-\ln\left( \phi b_k \right))  \right) \\
      &= - \phi \ln \phi \sum_k ( a_k + b_k) + \phi \sum_k ( a_k + b_k - a_k \ln a_k - b_k\ln b_k )
\end{align}
For the $\phi \ln \phi$ term the sum of eigenvalues is just the trace of the respective matrices 
which we can compute explicitly:
\begin{equation}
  \label{eq:alpha_int}
  \sum_k ( a_k + b_k) = \frac{1}{4\pi^2} \int_0^\pi dx \int_0^\pi dy \frac{ \sin(x) \sin(y)}{ \cos^3(\frac{x-y}{2}) ( \cos(\frac{x-y}{2})
    + \sin(\frac{x+y}{2}))} \approx .149
\end{equation}
It seems hard to find the coefficient of the term linear to $\phi$ using analytical method.
Instead we have from the direct eigenvalue decomposition of matrices $A$ and $B$ that
\begin{equation}
  \label{eq:beta_int}
  \sum_k ( a_k + b_k - a_k \ln a_k - b_k\ln b_k ) \approx .560
\end{equation}
These coefficients agree with our numerical results, see figure \ref{fig:Rferm_small_x}.

\section{The $x\to 1$ limit}

\label{sec:one}

We move on to studying the reflected entropy in the limit $x\to 1$ (or $\phi\to\pi/2$) in this section.
We will setup a systematic expansion for the eigenfunctions of the correlation matrix via a certain matching procedure between the endpoints of $A$.
Similar to QM scattering, solving for these matching conditions discretizes the eigenvalue spectrum and we obtain a finite reflected entropy.

\subsection{Reflected entropy}
\label{sec:rindler}
We need to reformulate our calculations adapted to $A$ being a half space cut. 
We pick $a_1 = 0, b_1 = \infty$ and $a_2 = -1/b, b_2 = -b$. The cross ratio of these points is:
\begin{equation}
  x = \frac{ (a_1-b_1)(a_2-b_2)}{(a_1 - a_2)(b_1-b_2)}
  =1 -b^2 
\end{equation}
where we restrict to $0 < b < 1$.  We also have:
\begin{equation}
  \label{full}
  Q =  \frac{z(z+1/b)(w_J+b)}{ w_J (z+b)(w_J+1/b)}
\end{equation}
where we now used directly $(z,w)$ (rather than $(x,y)$) as coordinates for the reflected fermion. Also $w_J = -w$. 
Taking the limit $b \rightarrow 0$ we find the correlator for the reflected fermion approaches (note that $Q \rightarrow 1$):
\begin{equation}
  \lim_{b \rightarrow 0} \left< \widetilde{\psi}^\dagger (z) \widetilde{\psi}(w) \right>_{\rho_{AA^c}}
  = \frac{\Theta_A(z) \Theta_A(w)+\Theta_{A^c}(z) \Theta_{A^c}(w) }{2\pi i (z-w-i \epsilon)} 
\end{equation}
The eigenfunctions of this operator are Rindler modes with a continuous spectrum. This continuum leads to a divergent entanglement (since the operator is no longer trace class).
These eigenfunctions are \cite{Casini:2009vk}:

\begin{equation}
  f_\nu^A(z) =\frac{1}{\sqrt{2\pi}} \theta(z)  z^{-1/2-i \nu} \qquad f_\nu^{A^c}(z) = \frac{1}{\sqrt{2\pi}}  \theta(-z)(-z)^{-1/2-i \nu}
\end{equation}
and these satisfy:
\begin{equation}
  \int_0^\infty dz \frac{1}{2\pi i (z-w - i \epsilon)} f_\nu^A(z)
  = \frac{1}{2}( 1-\tanh(\pi \nu))  f_\nu^A(w) \,, \qquad w \in A
\end{equation}
and
\begin{equation}
  \int_{-\infty}^0 dz \frac{1}{2\pi i (z-w - i \epsilon)} f_\nu^{A^c}(z)
  = \frac{1}{2}( 1+\tanh(\pi \nu))  f_\nu^{A^c}(w) \,, \qquad w \in A^c
\end{equation}

To resolve this continuum we need to take the limit $b \rightarrow 0$ more carefully. We hold $|z|/b, |w|/b$  fixed as we send $b \rightarrow 0$ (with fixed ratio $|z/w|$). This zooms in on the region near the entangling surface. Inside this region the correlator is:
\begin{equation}
  \delta \left< \widetilde{\psi}^\dagger (z) \widetilde{\psi}(w) \right>_{\rho_{AA^c}}
  \approx \frac{1}{2\pi i (z-w - 2 wz/b)} \Theta_{A}(z) \Theta_{A^c}(w) - (z \leftrightarrow w)
\end{equation}
Overall we can approximate the correlator as:
\begin{equation}
  C(z,w) \equiv \left< \widetilde{\psi}^\dagger (z) \widetilde{\psi}(w) \right>_{\rho_{AA^c}}
  \approx \frac{1}{2\pi i (z-w - 2 wz/b (\theta( -w)\theta(z) - \theta( w)\theta(-z) )}
\end{equation}
We can write this correlator as:
\begin{equation}
  C(z,w) =\left( \frac{ \partial \widetilde{z}}{\partial z} \right)^{1/2}
  \left(  \frac{ \partial \widetilde{w}}{\partial w} \right)^{1/2}
  \frac{1}{ 2\pi i(\widetilde{z} - \widetilde{w} - i \epsilon) }
\end{equation}
where
\begin{equation}
  \widetilde{z}(z) = z \theta(z) + \frac{z}{1-2z/b} \theta(-z)\,, \qquad \widetilde{w}(w) = w \theta(w) + \frac{w}{1-2w/b} \theta(-w)
\end{equation}
and the new coordinate satisfies $-b/2 \leq \widetilde{z} \leq \infty$. So the negative axis gets compactified. 
On this half space we know the Rindler eigenfunctions:
\begin{equation}
  \widetilde{f}_\nu (\widetilde{z}) = \frac{1}{\sqrt{2\pi}} \theta(\widetilde{z} + b) (\widetilde{z}+b)^{-1/2-i \nu}
\end{equation}
and these are the eigenfunctions of the integral Kernel $\propto (\widetilde{z} - \widetilde{w} - i \epsilon)^{-1}$. 
Taking into account the conformal scale factors the would be eigenfunctions of $C(z,w)$ are:
\begin{equation}
  \label{exp1}
  f_\nu(z) \approx \widetilde{f}_\nu (\widetilde{z}) \left(  \frac{ \partial \widetilde{z}}{\partial z} \right)^{1/2}
  = \frac{1}{\sqrt{2\pi}}\left( (z+b/2)^{-1/2- i \nu} \theta(z) + (b/2)^{-2 i \nu} (b/2-z)^{-1/2 + i \nu}  \theta(-z) \right)
\end{equation}
with eigenvalue $( 1-\tanh(\pi \nu))/2$. This eigenfunction behaves badly at $z \rightarrow \infty$.
This is to be expected since at this point (when $z \sim 1/b$) our scaling limit breaks down. 
Instead we must match onto a new solution here. If we again go back to the full correlator (determined by \eqref{full})
and hold fixed $|z| b$ and $|w| b$ as we send $ b \rightarrow 0$ then we find:
\begin{equation}
  \delta \left< \widetilde{\psi}^\dagger (z) \widetilde{\psi}(w) \right>_{\rho_{AA^c}}
  \approx \frac{1}{2\pi i (z-w + 2/b)} \Theta_{A}(z) \Theta_{A^c}(w) - (z \leftrightarrow w)
\end{equation}
Overall we can approximate the correlator here as:
\begin{equation}
D(z,w) \equiv \left< \widetilde{\psi}^\dagger (z) \widetilde{\psi}(w) \right>_{\rho_{AA^c}}
 \approx \frac{1}{2\pi i (z-w + 2/b (\theta( -w)\theta(z) - \theta( w)\theta(-z) )}
\end{equation}
which we can write as:
\begin{equation}
  D(z,w) =  \frac{1}{ 2\pi i(\widehat{z} - \widehat{w} - i \epsilon) }
\end{equation}
where
\begin{equation}
  \widehat{z}(z) = z \theta(z) + (z-2/b) \theta(-z)\,, \qquad \widehat{w}(w) = w \theta(w) + (w - 2/b) \theta(-w)
\end{equation}
the new coordinate satisfies $0 \leq \widehat{z} \leq \infty$ and $ -\infty \leq \widehat{z} \leq -2/b$ (which is a domain that wraps around $\infty$.)
This can be described via $1/\widehat{z} \geq -b/2$. 
On this domain we use the Rindler eigenfunctions:
\begin{equation}
  \widehat{f}_\nu (\widehat{z}) = \frac{1}{\sqrt{2\pi}}  \theta(1/\widehat{z} + b/2 ) (\widehat{z})^{-1} (1/\widehat{z} + b/2)^{-1/2+i \nu}
\end{equation}
which has the same eigenvalue, for the kernel $\propto (\widehat{z} - \widehat{w} + i \epsilon)^{-1}$ as we used in the previous patch.
The eigenfunction for the integral equation of interested, in this new scaling limit must then be:
\begin{equation}
  f_\nu(z) \approx \frac{1}{\sqrt{2\pi}} \left( \theta(z) z^{-1} (1/z+b/2)^{-1/2+ i \nu}
    + \theta(-z) z^{-1} (b/2)^{ 2 i \nu} (-1/z+b/2)^{-1/2 - i \nu} \right)
\end{equation}
The small $z$ limit of this expression should match onto the large $z$ limit of the previous expression \eqref{exp1}
up to an overall scaling:
\begin{align}
  \begin{split}
  \kappa &\left( \theta(z) z^{-1} (1/z)^{-1/2+ i \nu}
      + \theta(-z) z^{-1} (b/2)^{ 2 i \nu} (-1/z)^{-1/2 - i \nu} \right) \\
    &
    \approx \left( (z)^{-1/2- i \nu} \theta(z) + (b/2)^{-2 i \nu} (-z)^{-1/2 + i \nu}  \theta(-z) \right)        
  \end{split}
\end{align}
which gives $\kappa = 1$ and $ (b/2)^{4 i\nu} = -1$. This is our quantization condition.
More explicitly the two limits of the eigenfunction given above match smoothly if this
condition is satisfied. We have the following allowed values of $\nu$: 
\begin{equation}
  \nu_m  = \frac{\pi (m-\frac{1}{2})}{2 (-\ln(b/2))}
\end{equation}
where $m \in \mathbb{Z}$. These give rise to eigenvalues of $C$ as:
\begin{equation}
  \label{eq:lambdaleading}
  C_m = \frac{1 - \tanh(\pi \nu_m)}{2} = ( 1 + e^{2\pi \nu_m})^{-1}
\end{equation}
The reflected entropy is:
\begin{equation}
  \label{sum}
  S_R = \sum_{m=-\infty}^{\infty}( - C_m \ln C_m - (1-C_m) \ln (1-C_m)) 
\end{equation}
Since the spacing is small as $b \rightarrow 0$, we can approximate this by an integral with density:
\begin{equation}
  2 \int_0^1 d C  \left| \frac{d\nu}{dC} \frac{dm}{d\nu_m}  \right|  ( - C \ln C)
  = \frac{2 (-\ln(b/2))}{\pi^2} \int_0^1 d C  \frac{1}{ C(1-C)}   ( - C \ln C)
  = \frac{(-\ln(b/2))}{3}
\end{equation}
Writing this in terms of the cross-ratio we have:
\begin{equation}
  \label{eq:SR_leading}
  S_R \rightarrow \frac{1}{6} ( - \ln (1-x) + \ln 4 ) + \ldots
\end{equation}
which was as predicted by the replica trick \cite{Dutta:2019gen}. In principle we can compute corrections
to this as an expansion in $(1-x)$ and  $1/\ln(1-x)$.

In fact we can give a more general expression that is leading in $(1-x)$ but one that re-sums all $1/\ln(1-x)$ effects.
Firstly we notice that \eqref{sum} is the entropy deriving from a partition function:
\begin{equation}
  \label{eq:thermal_Z}
  S_R = - \beta^2 \partial_\beta \left( \beta^{-1} \ln Z(\beta) \right)  \qquad\, Z(\beta) = \prod_{m=0}^\infty (1 + q^{m+1/2})^2
\end{equation}
where $q = e^{i 2\pi \tau}$ and the inverse temperature is:
\begin{equation}
  \label{temp}
  \tau = i \frac{\pi}{2(- \ln( (1-x)/4))}
\end{equation}
We learn that the spectrum of the reflected density matrix approaches that of a free fermion
in the NS sector. So the reflected entropy is simply given by the appropriate Jacobi theta function.

And indeed, as is often the case, the universal term in \eqref{eq:SR_leading} arises simply via the Cardy formula. 
We would speculate that this is a more general result: the reflected entanglement spectrum
approaches that of the thermal partition function of the CFT under consideration, with temperature given by \eqref{temp}.
There seems to be some connection between the relfected entropy and the computable cross norm negativity \cite{Yin:2022toc} (at least for some Renyi generalization
of reflected entropy) which could possibly be used to give a more general proof of this fact. 

\subsection{Deflected entropy}
\label{sec:deflected}

We now study a generalization of the reflected entropy that we call deflected entropy.
It is given by applying a modular flow on the correlation function \eqref{eq:correlator} by an additional amount of $s \in \mathbb{R}$:
\begin{align}
  \Braket{\tilde{\psi}^\dag(x)\tilde{\psi}(y)}_{\rho_{AA^c}} = 
  \begin{cases}
    \Braket{\psi^\dag(x) \psi(y)}_\Omega, \quad &x,y \in A \,\, {\rm or } \,\, x,y \in A^c \\
    \left( \frac{\partial y_J}{\partial y} \right)^{1/2}\Braket{\psi^\dag(x) \Delta^{1/2+is}_{AB} \psi(y_J)}_\Omega, &x\in A, y\in A^c \\
     \left( \frac{\partial x_J}{\partial x}\right)^{1/2} \Braket{\psi^\dag(x_J) \Delta^{1/2-is}_{AB} \psi(y)}_\Omega, &y\in A, x\in A^c
  \end{cases}
\end{align}

When $s\neq0$ this correlator is not continuous across the entangling surface.
To fix this discontinuity we replace the geometric reflection map by another vacuum modular flow (with respect to region $A$) with an amount $-s+i/2$.
In our current coordinate settings this action is simply
\begin{align}
  x\to x_S = -e^{2\pi s}x
\end{align}
This extra vacuum modular flow for the A region leaves the entropy invariant since it is generated by an operator that acts solely in $AA^\star$ (working in the original frame of $AB A^\star B^\star$) - the easiest way to see this is by using the modular operator for the split state, which agrees with the vacuum flow inside A.
When $s=0$ we simply get back the geometric reflection; whereas when $s\neq 0$ this additional flow smooths out the jump discontinuity across the entanglement surface.
Algebraically one can think of this construction as performing a Connes cocycle flow \cite{Levine:2020upy, Ceyhan:2018zfg, ASENS_1973_4_6_2_133_0} defined with the $AB$ 
algebra and for the vacuum and  split state. See for example \cite{Levine:2020upy}.

The analysis of the previous subsection still carries over as long as we assume $b < e^{-\pi s}$ when we scale $b\to 0$.
We can approximate the new correlation function as
\begin{align}
  C(x,y) \simeq
  \begin{cases}
    \dfrac{1}{2\pi i} \dfrac{1}{x-y-\frac{2xy}{b}e^{\pi s}\cosh(\pi s)(\theta(x)\theta(-y)-\theta(-x)\theta(y))},\quad |x|,|y|\ll e^{-\pi s}\\
    \dfrac{1}{2\pi i} \dfrac{1}{x-y+\frac{2}{b}e^{-\pi s}\cosh(\pi s)(\theta(x)\theta(-y)-\theta(-x)\theta(y))},\quad |x|,|y|\gg e^{-\pi s}\\    
  \end{cases}
\end{align}
The effect of the Connes cocyle flow is to shift $b$ in both regimes accordingly and the matching point from $1$ to $e^{-\pi s}$.
One can show the spectrum is now determined by:
\begin{equation}
  \nu_m = \frac{\pi(m-1/2)}{2\ln (2\cosh (\pi s)/b)}
\end{equation}
such that the reflected entropy becomes:
\begin{equation}
\label{tocompff}
  S_R \rightarrow \frac{1}{6}\left( - \ln (1-x) + \ln 4 + 2\ln (\cosh(\pi s )) \right) + \ldots
\end{equation}
This result also agrees well with our numerics, see Figure~\ref{fig:s_plots}.

In the Appendix~\ref{app:A} we perform a different computation of the deflected entropy
in AdS$_3$/CFT$_2$. We use this to compare to the free fermion computations. We find:
\begin{align}
  S_{R,\rm holographic} = \frac{c}{6} \ln \left| \frac{1+\sqrt{w}}{1-\sqrt{w}}  \right|+ O(c^0), \quad w = \frac{x(1+e^{2\pi s})^2}{(x+e^{2\pi s})(1+xe^{2\pi s})}
\end{align}
From which we get the following asymptotic when $x\to 1$:
\begin{align}
  S_{R,\rm holographic} = \frac{c}{6} (-\ln(1-x)+\ln 4 + \ln(\cosh(\pi s))) + \cdots
\end{align}
The overall factor of $c$ (where in holographic theories $c$ is large) accounts for the extra degrees of freedom. 
Comparing to \eqref{tocompff} we see similar behavior aside from the factor of $2$. There is of course no reason to expect agreement.
This difference persists away from $x \rightarrow 1$ as can be seen in Figure~\ref{fig:s_plots}. 

\subsection{Next order corrections}
\label{sec:corrections}

In this subsection we systematically improve on our results for the spectrum as $x \rightarrow 1$.
This also serves to convince the reader that our results are under control.

We firstly setup the leading order answer more carefully. We will construct approximate eigenfunctions of the integral equation,
valid as $b \rightarrow 0$ (that is $x \rightarrow 1$). 
This also  fills in some holes in the original discussion
in Section~\ref{sec:rindler}. 
This subsection is a bit technical so the reader might prefer to skip to the conclusions.

Let us see that the functions we wrote down are really approximate eigenfunctions as $b \rightarrow 0$. 
\emph{Define} the function:
\begin{align}
  g_{\nu}(z) &\equiv \nonumber \left( (z+b/2)^{-1/2- i \nu} \theta(z) + (b/2)^{-2 i \nu} (b/2-z)^{-1/2 + i \nu}  \theta(-z) \right)
             \theta(1- |z |)
  \\& + \left( \theta(z) z^{-1} (1/z+b/2)^{-1/2+ i \nu}
  + \theta(-z) z^{-1} (b/2)^{ 2 i \nu} (-1/z+b/2)^{-1/2 - i \nu} \right) \theta(|z |-1) 
\end{align}
where we have designated an arbitrary matching point of $|z|=1$. 
This function has a non-uniform expansion as $b \rightarrow 0$. For $|z| \sim b$ then $|g_\nu(z)| \sim b^{-1/2}$,
for $|z| \sim 1/b$ then $|g_\nu(z)| \sim b^{1/2}$ and finally for $|z| \sim 1$ then $g_\nu(z) \sim 1$. 

Consider, for $w>0$ the following integral:
\begin{align}
  \int_{-\infty}^{\infty} d z C(z,w) g_{\nu}(z)
\end{align}
We need to check this integral in the various regimes. We first consider $ w \sim b$. 
One can see that the integral is dominated by the regime $-1/w < z< 1/w$ giving an answer $\mathcal{O}(b^{-1/2})$:
\begin{align}
  \int_{-\infty}^{\infty} d z C(z,w) g_{\nu}(z)
  \approx \lambda_\nu g_{\nu}(w) 
\end{align}
Similarly for $w \sim 1/b$ the integral is dominated in the region $ z>1/w$ and $z< -1/w$ giving an answer $\mathcal{O}(b^{1/2})$ with the same form as above.
Finally for  $ w \sim 1$ we only get a contribution from away from the scaling regimes, and $z>0$ (since the kernel is otherwise suppressed by $\mathcal{O}(b)$ for $z<0$ when $w>0$) where:
\begin{equation}
  \int_{-\infty}^{\infty} d z C(z,w) g_{\nu}(z) \approx \int_0^\infty dz \frac{1}{2\pi i(z-w - i\epsilon)} z^{-1/2 - i\nu} 
  = \lambda_\nu  w^{-1/2 - i\nu} \approx \lambda_\nu g_\nu(w)
\end{equation}
Thus, at least for $w>0$ it looks like $g_\nu$ is an approximate eigenfunction with eigenvalue $\lambda_\nu$. 

A similar analysis applies to the case $w<0$, where the $\mathcal{O}(1)$ regime is now:
\begin{align} \nonumber
  &\int_{-\infty}^{\infty} d z C(z,w) g_{\nu}(z)  \approx \int_{-\infty}^0 dz \frac{ (-z)^{-1/2 + i \nu} }{2\pi i(z-w - i\epsilon)}  \left(  (b/2)^{-2 i \nu}\theta(1+z)
    - (b/2)^{ 2 i \nu} \theta(-(z+1))  \right)
  \\ & = - \lambda_\nu (b/2)^{2 i \nu} (-w)^{-1/2 + i \nu} +  ((b/2)^{-2 i \nu} + (b/2)^{2 i \nu} )\int_{-1}^0 dz \frac{ (-z)^{-1/2 + i \nu} }{2\pi i(z-w - i\epsilon)}  \\
  & \approx \lambda_\nu g_\nu(w) + [ (b/2)^{-2 i \nu} + (b/2)^{2 i \nu}] \left(- \lambda_\nu \theta(1 + w) (-w)^{-1/2 + i \nu}+ \int_{-1}^0 dz \frac{ (-z)^{-1/2 + i \nu} }{2\pi i(z-w - i\epsilon)}  \right)
\end{align}
So all together:
\begin{equation}
  \int_{-\infty}^\infty dz (C(z,w) - \lambda_\nu \delta(z-w)) g_\nu(z) \approx [ (b/2)^{-2 i \nu} + (b/2)^{2 i \nu}]  G_\nu(w)
\end{equation}
where $G_\nu(w)$ is $\theta(-w)$ times the function above and satisfies
$G_\nu(w)  \sim \mathcal{O}(1)$ if $|w| \sim 1$ and $\sim \mathcal{O}(1)$ for $|w| \sim b$
and  $\sim \mathcal{O}(b^{1})$ if $|w| \sim 1/b$. Thus we find that $g_\nu$ is only an approximate eigenfunction upon imposing the
quantization condition discussed in the previous subsection. 

We can now prove approximate orthogonality for distinct eigenvalues:
\begin{equation}
  0 \approx \int_{-\infty}^{\infty} dw \int_{-\infty}^\infty dz g_{\nu_m}^\star(w) (C(z,w) - \lambda_{\nu_n} \delta(z-w)) g_{\nu_n}(z) 
  \approx ( \lambda_{\nu_m}- \lambda_{\nu_n}) \int_{-\infty}^{\infty} dw g_{\nu_m}^\star(w)  g_{\nu_n}(w) 
\end{equation}
and we can compute the normalization by moving slightly away from the quantization condition for one of the $\nu$'s above:
\begin{equation}
  ( \lambda_{\nu_m}- \lambda_{\nu}) \int_{-\infty}^{\infty} dw g_{\nu_m}^\star(w)  g_{\nu}(w) 
  \approx  [ (b/2)^{-2 i \nu} + (b/2)^{2 i \nu}] \int_{-\infty}^0 dw g_{\nu_m}^\star(w) G_\nu(w) 
\end{equation}
One can see that the later integral $I = \int_{-\infty}^0 dw g_{\nu_m}^\star(w) G_\nu(w)$ is dominated in the order one regime where it evaluates to:
\begin{align}
  I =  e^{i\pi(m-1/2)} \int_{-\infty}^0 dw \left(- \lambda_{\nu} \theta(1 + w) (-w)^{-1/2 + i \nu}+ \int_{-1}^0 dz \frac{ (-z)^{-1/2 + i \nu} }{2\pi i(z-w - i\epsilon)}  \right) (-w)^{-1/2 - i \nu_m}  
\end{align}
This integral is most efficiently evaluated by taking $\nu$ complex and so that the $w$ integral can be done first (the integral is well defined as stated above)
\begin{align}
  I = e^{i\pi(m-1/2)}\left( \frac{\lambda_{\nu}- \lambda_{\nu_m}} {i( \nu-\nu_m)} \right)  
\end{align}
Thus:
\begin{equation}
  \int_{-\infty}^{\infty} dw g_{\nu_m}^\star(w)  g_{\nu_n}(w) \approx \delta_{m,n} 4 (- \ln(b/2) )
\end{equation}
This gives (at least formally) the completeness relation:
\begin{equation}
  \label{complete}
  \sum_{m=-\infty}^{\infty} g_{\nu_m}^\star(w) g_{\nu_m}(w') \approx \frac{1}{4 (- \ln(b/2) )} \delta(w-w')
\end{equation}

We can now attempt to find corrections to these eigenfunctions and eigenvalues using perturbation theory. 
We write:
\begin{equation}
  h_m(z) = g_m(z)  
  +  \sum_{n=-\infty}^{\infty} \delta \mu_{mn} g_n(z)  + \ldots
\end{equation}
where $\delta \mu_{mn} = \mathcal{O}(b)$.  Plugging this into the eigenfunction equation:
\begin{equation}
  \int_{-\infty}^{\infty} dz C(z,w) h_m(z) = (\lambda_m + \delta \lambda_m) h_m(w)
\end{equation}
To first order in $b$ we have:
\begin{align}
  \label{tolead}
  \begin{split}
      4 (-\ln(b/2)) \delta \lambda_m &= \int_{-\infty}^{\infty} dw\left(   -\lambda_m \delta(g_m(w))+ \delta \int_{-\infty}^{\infty} dz C(z,w) g_m(z)
                        \right)g_m^\star(w) \\
                        &= \int_{-\infty}^{\infty} dw \int_{-\infty}^{\infty} dz\delta( C(z,w) ) g_m(z) g_m^\star(w)
  \end{split}
\end{align}
which does not depend on $d_{mn}$ or $\delta \nu_m$. One sees the usual simplicity of 
first order perturbation theory.  

We now seek the leading term in \eqref{tolead}.
We expect an expansion of the form:
\begin{equation}
  \delta\lambda_m = \sum_{k=1}^\infty b^{k} \ell_k( \ln b)
\end{equation}
where $\ell_k$ are  functions of $\ln b$.
We will aim for the order $k=1$ term. 

Since the integrand in \eqref{tolead} has a non-uniform expansion in $b$ we must expand it in various regions
$z,w \sim b, 1, b^{-1}$. We expand only in $C(z,w)$, keeping the leading terms in the wave-functions and the measure $dz dw$ (which gets $b$ dependence after
scaling into these regions.)
This procedure goes under the name of matched asymptotic expansion \cite{Bender1999, Faulkner:2008hm, Faulkner:2009wj}. One often finds power law divergences
after isolating the various regions using scaling arguments. The general rule is that, since the total integral is well defined, then any power law divergences
will cancel amongst the various regions, that is after a re-arrangement of the order in the $b$ expansion (these divergences
typically lead to enhancements in the $b$ expansion that mix between the orders.)  In some instances we can proceed by using dimensional regularization. We give $\nu$
a small imaginary part and this removes any power law diverges (log's would show up as poles, but we do not find any.)

The diagonal regions $(z\sim 1,w \sim 1)$,  $(z\sim b,w \sim b)$ and $(z\sim 1/b,w \sim 1/b)$ all give rise to $\mathcal{O}(b^2)$ corrections.
Since the pole/singularity in the kernel lives in the diagonal regions, we have the symmetry $\delta (C(z,w)) = - \delta ( C(w,z)) = \delta ( C(w,z))^\star$
which allows us to restrict the integral to $z > w$:
\begin{align}
  4 (-\ln(b/2)) \delta \lambda_m  &= \int_{z>w} dw dz\delta( C(z,w) ) g_m(z) g_m^\star(w)
                         +  \int_{z<w} dw dz\delta( C(z,w) ) g_m(z) g_m^\star(w) \nonumber\\ &
                                                                          = 2 {\rm Re} \int_{z>w} dw dz\delta( C(z,w) ) g_m(z) g_m^\star(w)
\end{align}
after re-labeling $w \leftrightarrow z$ in the second term.

The non-diagonal regions with one $z\sim 1$ or $w \sim 1$ naively give $b^{1/2}$ contribution, but they evaluate to simple power laws that do not contribute after following
the above rules. We are left with the crossed regions $(z\sim b, w\sim 1/b)$ and $(z\sim 1/b, w\sim b)$.
Note that the correlator satisfies $C(z,w) = C(z^{-1},w^{-1})/(zw)$ under inversion.
Also $g_m(1/z) = g^*_m(z)/z$, one finds that the contribution from the two crossed regions are equal. \footnote{This follows from the scaling behavior of the 2D free Fermions.}
Therefore we only need to study one crossed region, say $(z\sim 1/b, w\sim b)$:
\begin{align}
  \begin{split}
  \delta C(z,w) &=
  \frac{1}{2 \pi i z}\left( \theta(zw)  + \frac{  (w/b+1)( z b-1) + w/b}{ (w/b+1)( z b-1) - w/b} \theta(-z) \theta(w) \right. \\
  & + \left. \frac{  (w/b-1)( z b+1) - z/b}{ (w/b-1)( z b+1) + z/b} \theta(-z) \theta(w) \right)
\end{split}
\end{align}
The wave function in this region is
\begin{align}
  g_{m}(z) g_{m}^\star(w) &\equiv \nonumber
                        i (-)^m \left(  (1+2w/b)^{-1/2+ i \nu_m}\theta(w) +    (1-2w/b)^{-1/2 - i \nu_m} \theta(-w) \right)
  \\&  \times \left(  (2/(zb))   (1+2/(zb))^{-1/2+i \nu_m} \theta(z)
  +    (2/(zb))   (1-2/(zb))^{-1/2 - i \nu_m} \theta(-z)\right)
\end{align}
It is convenient to use the scaled coordinates (related to $\hat{z}, \widetilde{z}$ defined above):
\begin{equation}
  x = (1+2/(bz)) 
  \,, \qquad 
  y = (1+2w/b) \theta(w) +  (1- 2w/b))^{-1} \theta(-w) 
\end{equation}
where we have assumed $z > 0$. In the scaled region we cover $0 < y< \infty$ and $1 < x < \infty$.
After rescaling to $x,y$ coordinates we find:
\begin{align}
  \begin{split}    
4 (-\ln(b/2)) \delta \lambda_m  
= \frac{ b (-)^m}{\pi} {\rm Re} & \left( \int_{1}^{\infty} d x x^{-1/2+i \nu_m} \int_0^1 d y y^{-1/2+i \nu_m} \frac{ 1+ x y}{y (x+y)}  \right.
\\ & \qquad  + \left. \int_{1}^{\infty} d x x^{-1/2+i \nu_m} \int_1^\infty d y y^{-1/2+i \nu_m} \right)
  \end{split}
\end{align}
These later integrals converge if we analytically continue the various appearances of $\nu_m$ differently. While this is rather
crude, it does the job of extract the non-power law divergences. 

\begin{figure}[h]
  \centering
  \includegraphics[width=.6\textwidth]{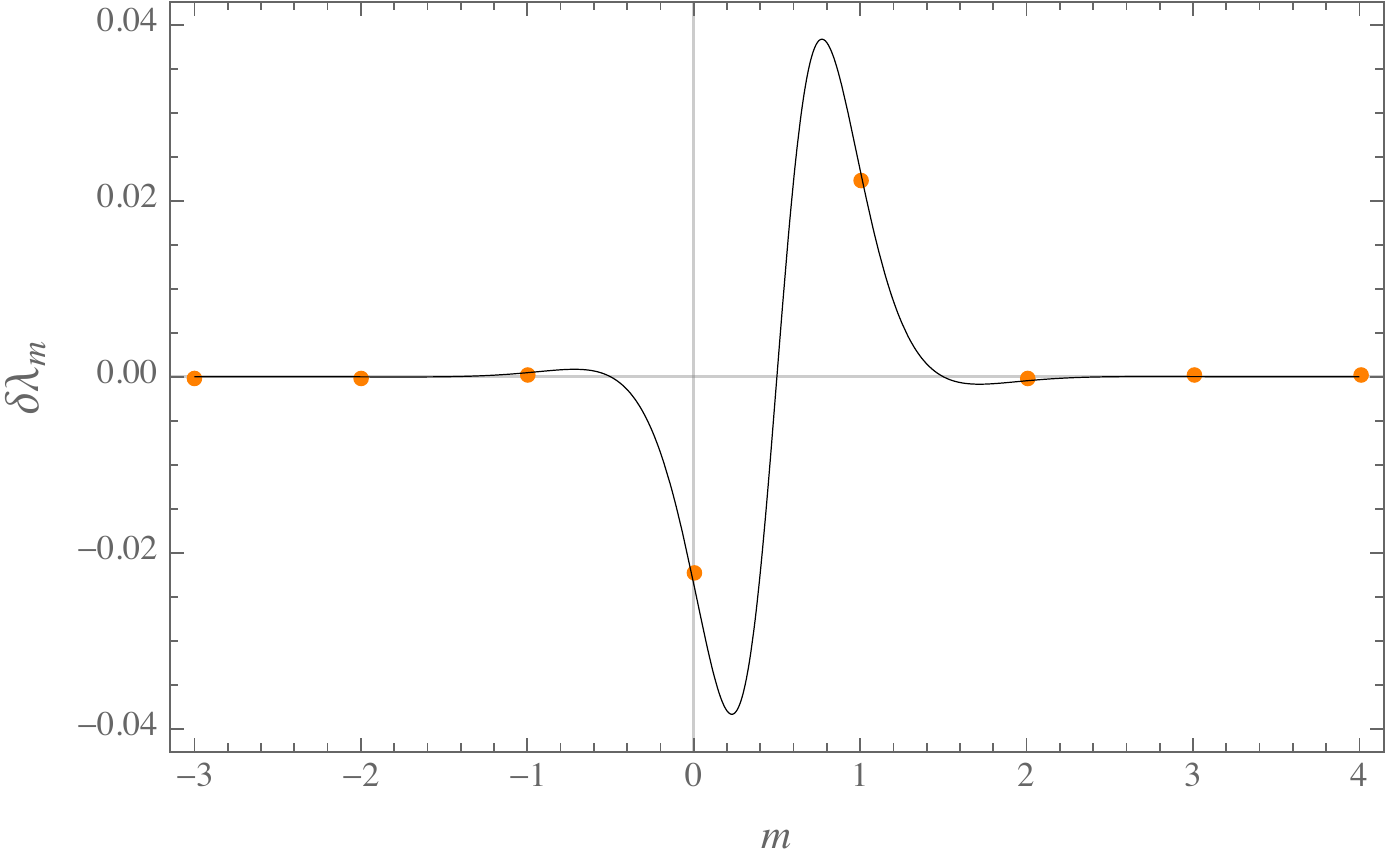}
    \includegraphics[width=.6\textwidth]{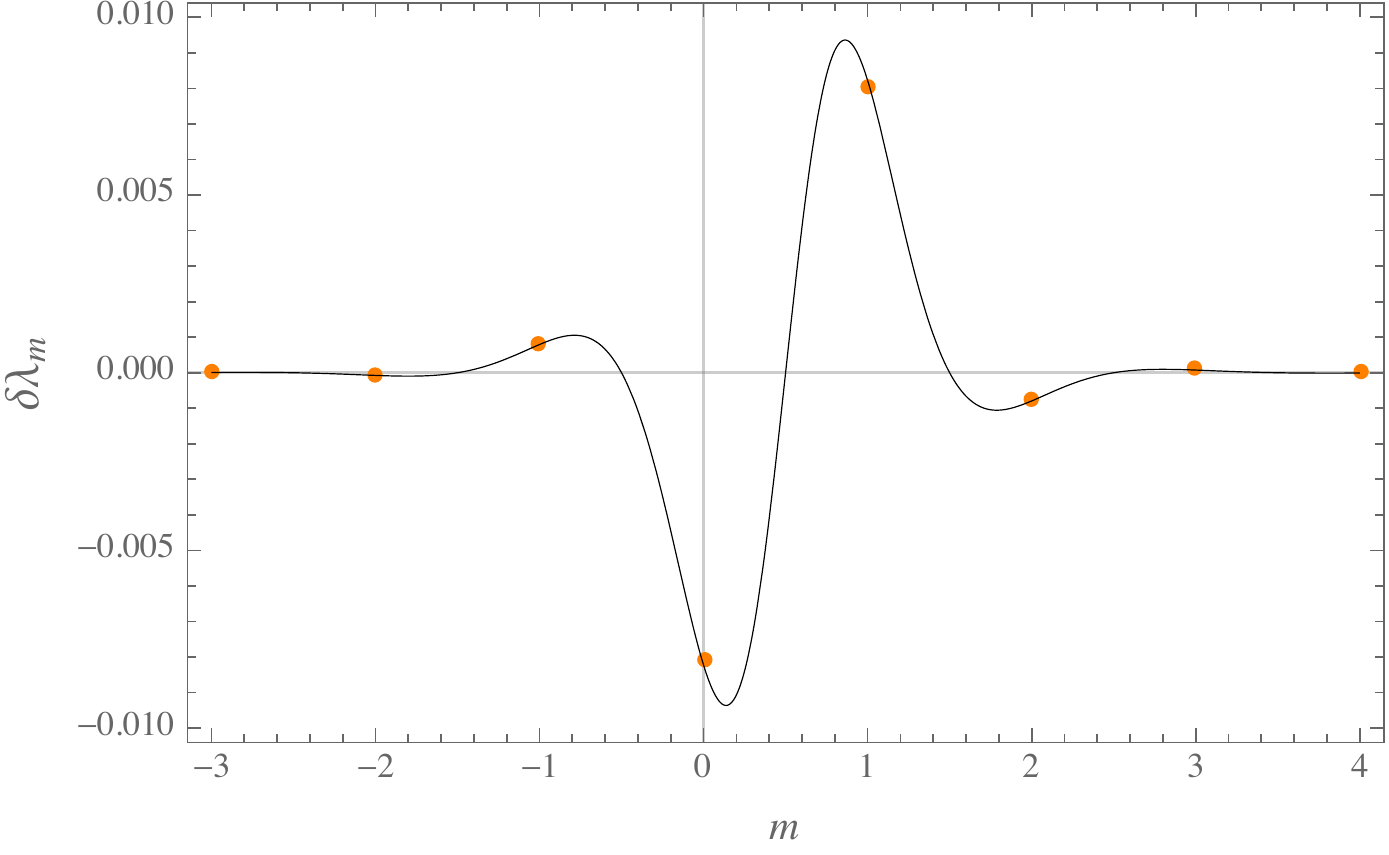}
    \caption{The difference $\delta\lambda_m$ between the numerical eigenvalues and the leading order analytics, plotted with the predictions from next order corrections  \eqref{eq:lambdashift}. The cross-ratio in these figures are $x=0.914$ (top) and $0.9987$ (bottom), respectively.}
    \label{fig:lambdashift}
\end{figure}

The final result is:
\begin{equation}
  \label{eq:lambdashift}
\delta \lambda_m = \frac{b}{2 ( - \ln b/2)}  \frac{(-)^m \sech(\pi \nu_m)}{(1+ 4\nu_m^2)} + \mathcal{O}(b^2) 
\end{equation}
This leads to a shift in the inferred spectrum:
\begin{equation}
\delta m = \frac{b}{2}   \frac{(-)^m \cosh(\pi \nu_m)}{\pi^2(1/4+ \nu_m^2)} + \ldots
\end{equation}
Which competes with the leading order answer
when $  m \sim \frac{2}{\pi^2} (-\ln b/2)^2$.  While we don't expect perturbation theory to break down here (since $\delta \lambda_m$ remains small), it will become
difficult to extract the spectrum at this order of $m$, since the density matrix has an exponentially small depends on these energy shifts.
These shifts that we found analytically agrees surprisingly well with our numerics, see Figure~\ref{fig:lambdashift}:

With the eigenvalue corrections at hand, one can check that the correlation matrix (hence also the density matrix) converges to the thermal answer \eqref{eq:lambdaleading} by examining the distance between two matrices.
Given spectrum of two operators one can define the \textit{spectral distance} by minimizing over the sum of the differences between two set of eigenvalues.
The spectral distance is known to be equivalent to the distance between unitary orbits of two operators \cite{Azoff84}.
In our setting this is simply
\begin{equation}
  \sum_{m} |\delta\lambda_m| = \sum_m\frac{b}{2(-\ln b/2)} \frac{\sech (\pi \nu_m)}{(1+4\nu^2_m)}
\end{equation}
It is easy to see that the sum is upper bounded by
\begin{align}
  \begin{split}    
  \sum_{m} |\delta\lambda_m| \le \frac{b}{(-\ln b/2)} \left(\int^\infty_{1/2} \sech(\pi \nu_m)dm + \sech(\pi\nu_1)\right) = 2\pi b + \O (b/\ln(b))
  \end{split}
\end{align}
which approaches zero as $b\to 0$.

\begin{figure}[h]
  \centering
  \includegraphics[width=.7\textwidth]{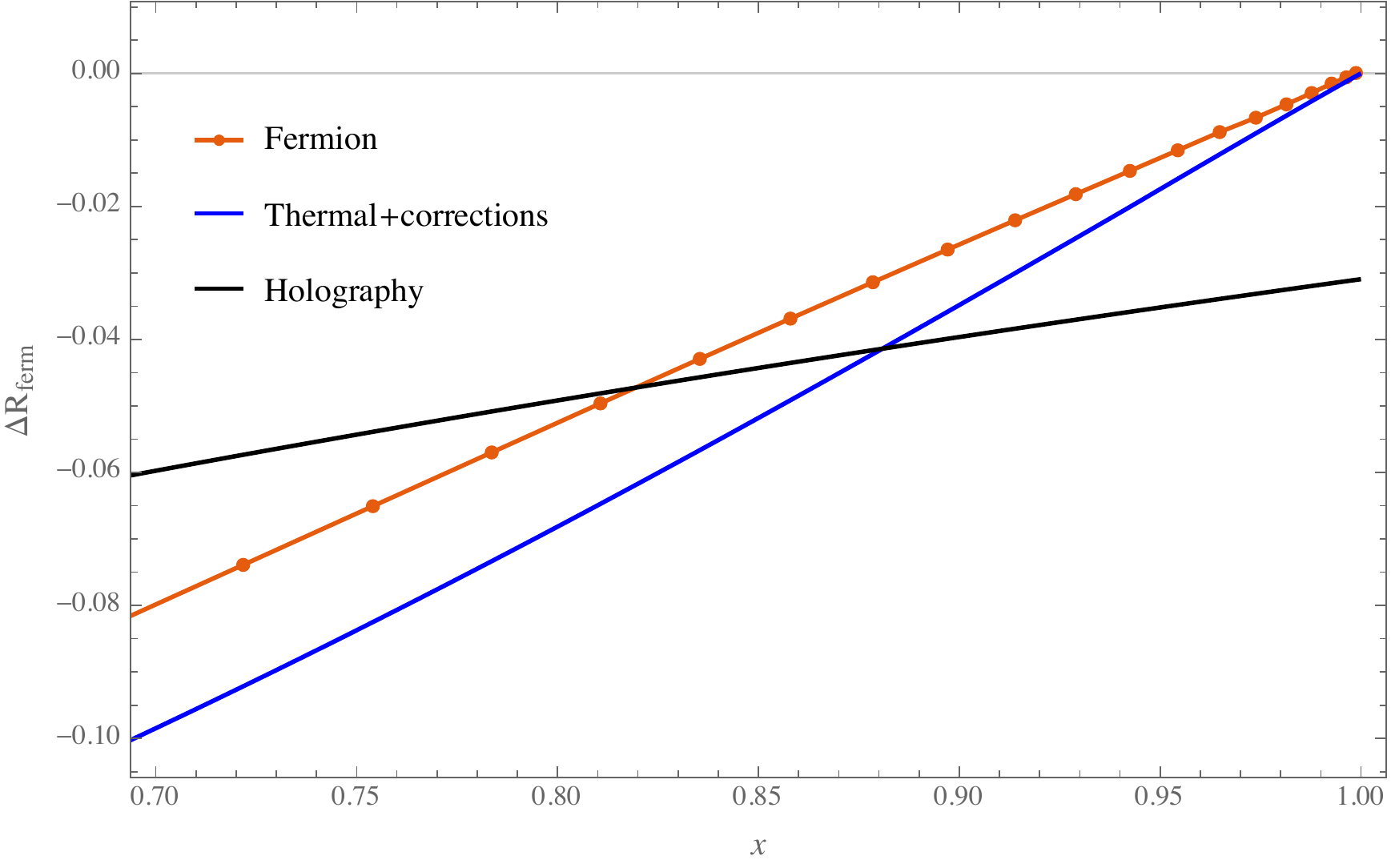}
  \includegraphics[width=.7\textwidth]{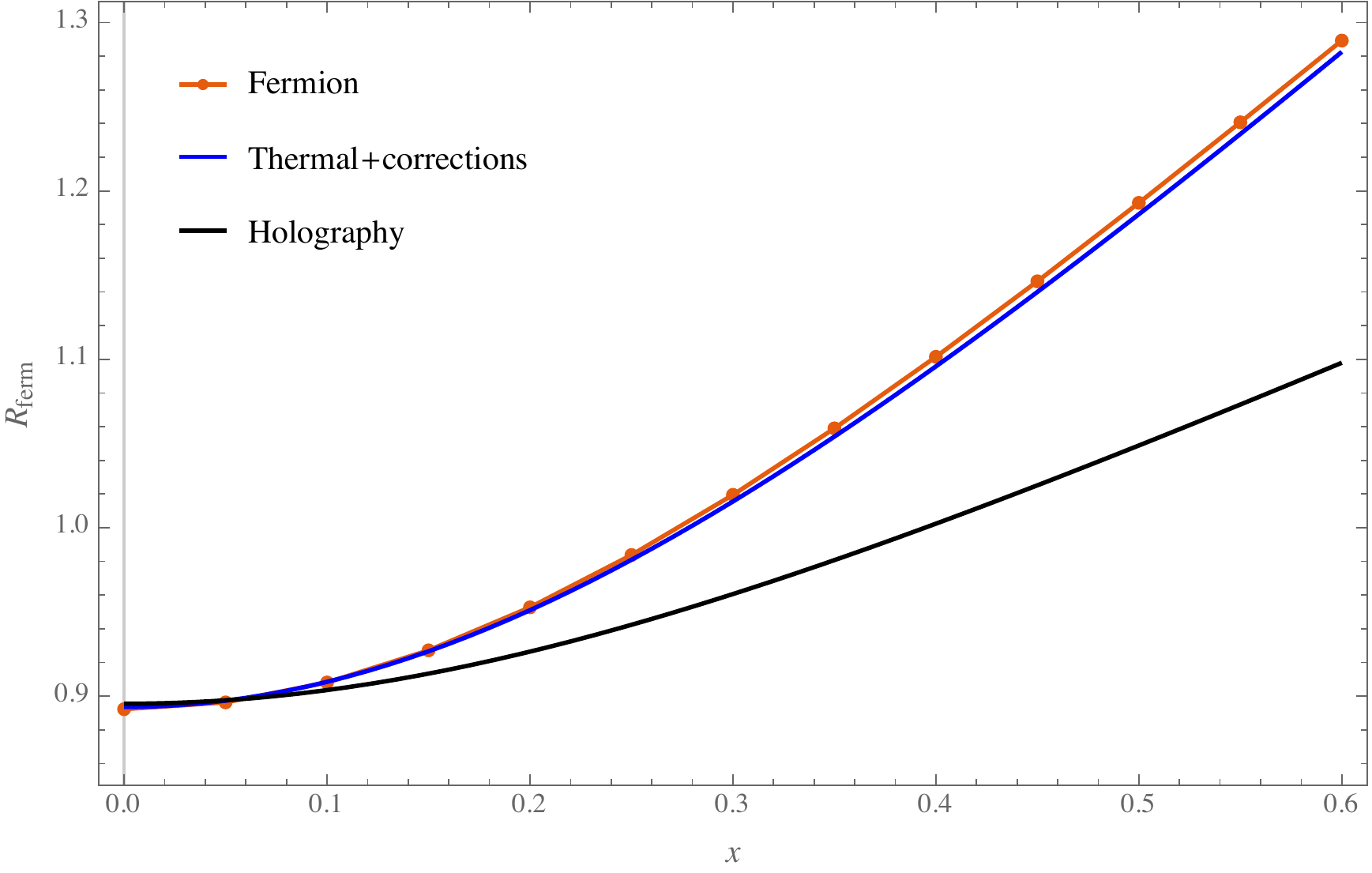}
  \caption{(Top) The free fermion deflected entropy $\Delta R_{\text{ferm}}$ for fixed $s=0.2$  with the asymptotic term $1/6(-\ln(1-x)+2\ln(\cosh{\pi s}) )$ subtracted, plotted against the thermal prediction with corrections \eqref{eq:s_lambdashift} and holographic CFTs.
    (Bottom) The deflected entropy for fixed cross-ratio $x=0.981$ and variable $s$, plotted against the same analytical prediction and holographic CFTs.}
  \label{fig:s_plots}
\end{figure}

Repeating the same analysis for deflected entropy correlator $C^s(z,w)$ with non-zero $s$, one gets (For details see Appendix~\ref{app:B}):
\begin{align}
  \label{eq:s_lambdashift}
  \delta\lambda^s_m = \frac{b^s}{2(-\ln b^s/2)}\frac{(-)^m\sech(\pi \nu^s_m)}{(1+4\nu^{s}_m)^2} \left( \cosh(\pi s)\cos(2\pi s \nu^s_m) - \frac{\sinh(\pi s)\sin(2\pi s\nu^s_m)}{2\nu^s_m} \right) + \ldots
\end{align}
where $b^s = b/\cosh(\pi s)$ and $\nu^s_m = \frac{\pi (m-1/2)}{-2\ln(b^s/2)}$ is the effective spectrum parameters for $s\ne0$.
We see that it reduces to the unflowed case when $s=0$.
It gives a prediction of deflected entropy that agrees well with our numerics (Figure~\ref{fig:s_plots}).

\section{Discussion}

\label{sec:discussion}

In this note we calculated the reflected entropy of a chiral free fermion both numerically and analytically, and we give asymptotic formulas for $S_R$ in both the limits $x\to 0$ and $x\to1$.
In particular we have shown that in the latter case, the reflected spectrum of a free chiral fermion in the NS sector approaches a thermal spectrum with inverse temperature to length ratio $\beta/L \propto -1/\ln(1-x)$.
Indeed this temperature coincides with the expected answer one would have obtained in AdS/CFT.
This thermal spectrum is theory dependent, as opposed to the universal spectrum of entanglement entropy in 2d CFTs \cite{Calabrese08} which one can regard as being in the infinite temperature limit.

As a practical matter one can then use the reflected entropy to extract the spectrum of operator dimensions directly from the vacuum wave-function.
In this regard, the reflected entropy can be used as a diagnostic of different topological orders \cite{Kudler-Flam:2020url}.
However, due to the perturbative corrections and Boltzman suppression,  the weights extracted in this way are only perturbatively reliable for a window of operator weights satisfying:
\begin{equation}
\label{mod}
h \lesssim \frac{1}{2\pi^2}  ( \ln (1-x) )^2 
\end{equation}
It is natural to conjecture that the above statement are more general statements about the reflected entanglement spectrum for any 2d CFT, with likely some modification of \eqref{mod} to include a varying central charge.

\acknowledgments

This work is partially supported by the Air Force Office of Scientific Research under award number FA9550-19-1-0360 and the Department of Energy under award number DE-SC0019183.

\appendix

\section{Deflected entropy for holographic CFTs}
\label{app:A}
In this appendix we present a quick derivation for the holographic deflected entropy using a replica trick.
We work in the same Rindler coordinate as in Section~\ref{sec:rindler}.
Consider the vacuum reduced density matrix $\rho_{AB}$ where we define $A =[a_1,b_1]$ and $B = [a_2,b_2]$.
The deflected entropy $R_s$ for this state is defined by the entanglement entropy of the following density matrix
\begin{align}
  \rho_{AA^*} = \text{Tr}_{BB^*} \ket{\rho^{1/2+is}_{AB}}\bra{\rho^{1/2+is}_{AB}}
\end{align}
We will construct this density matrix by first calculating the holographic entanglement entropy of the following replica state
\footnote{The change of sign from $is\to -m$ in the bra comes from the fact $\bra{\rho^{n/2+is}} = \bra{\rho^{n/2}}\rho^{-is}$.}
\begin{align}
  \rho^{m,n}_{AA^*} = \text{Tr}_{BB^*}\ket{\rho^{n/2+m}}\bra{\rho^{n/2-m}}, \quad n \in 2\mathbb{Z}_+, m \in \mathbb{Z}
\end{align}
and then take the analytic continuation $n\to1,m\to is$.

\begin{figure}[h]
  \centering
  \includegraphics[width=.45\textwidth]{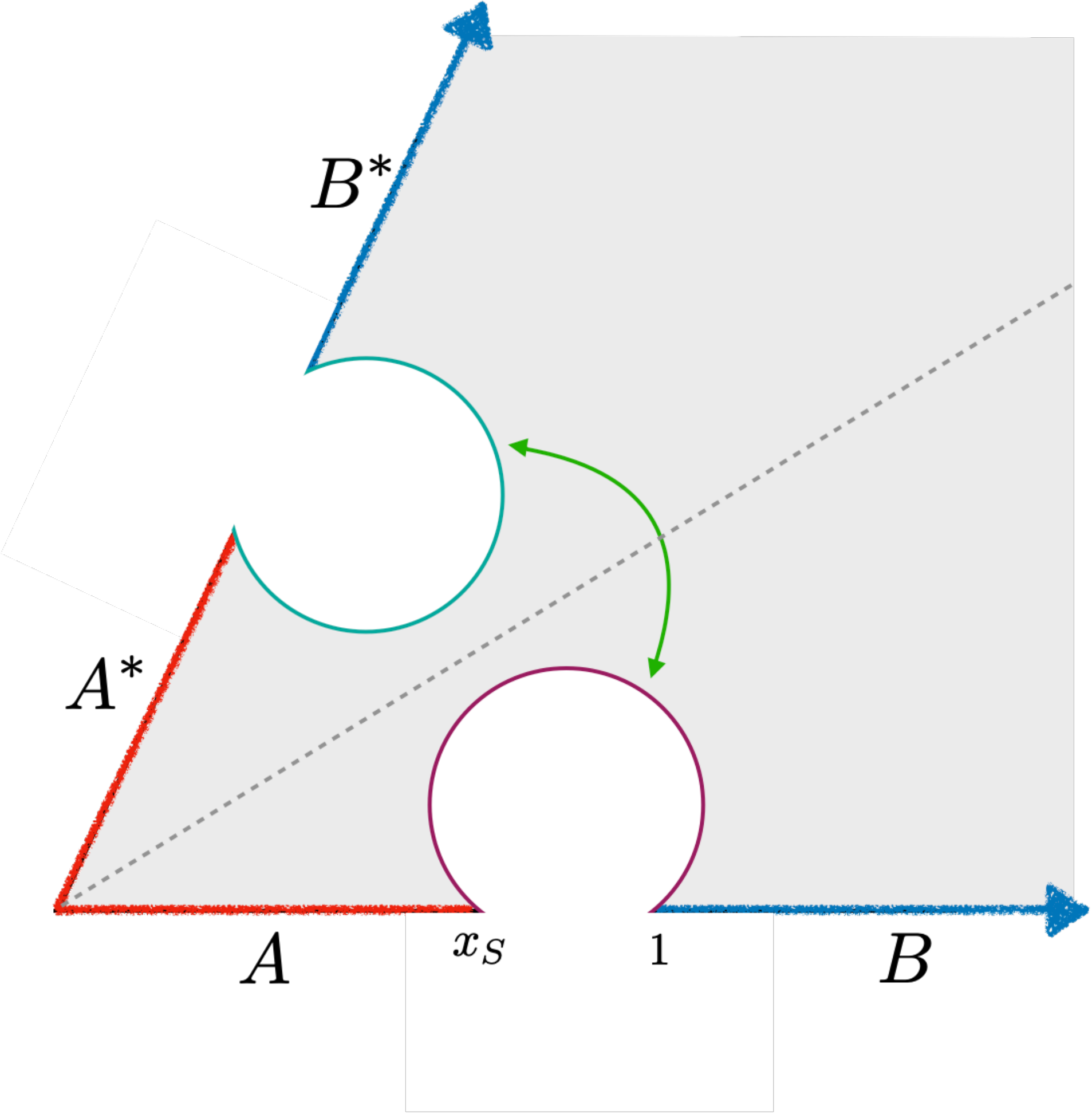}
  \includegraphics[width=.54\textwidth]{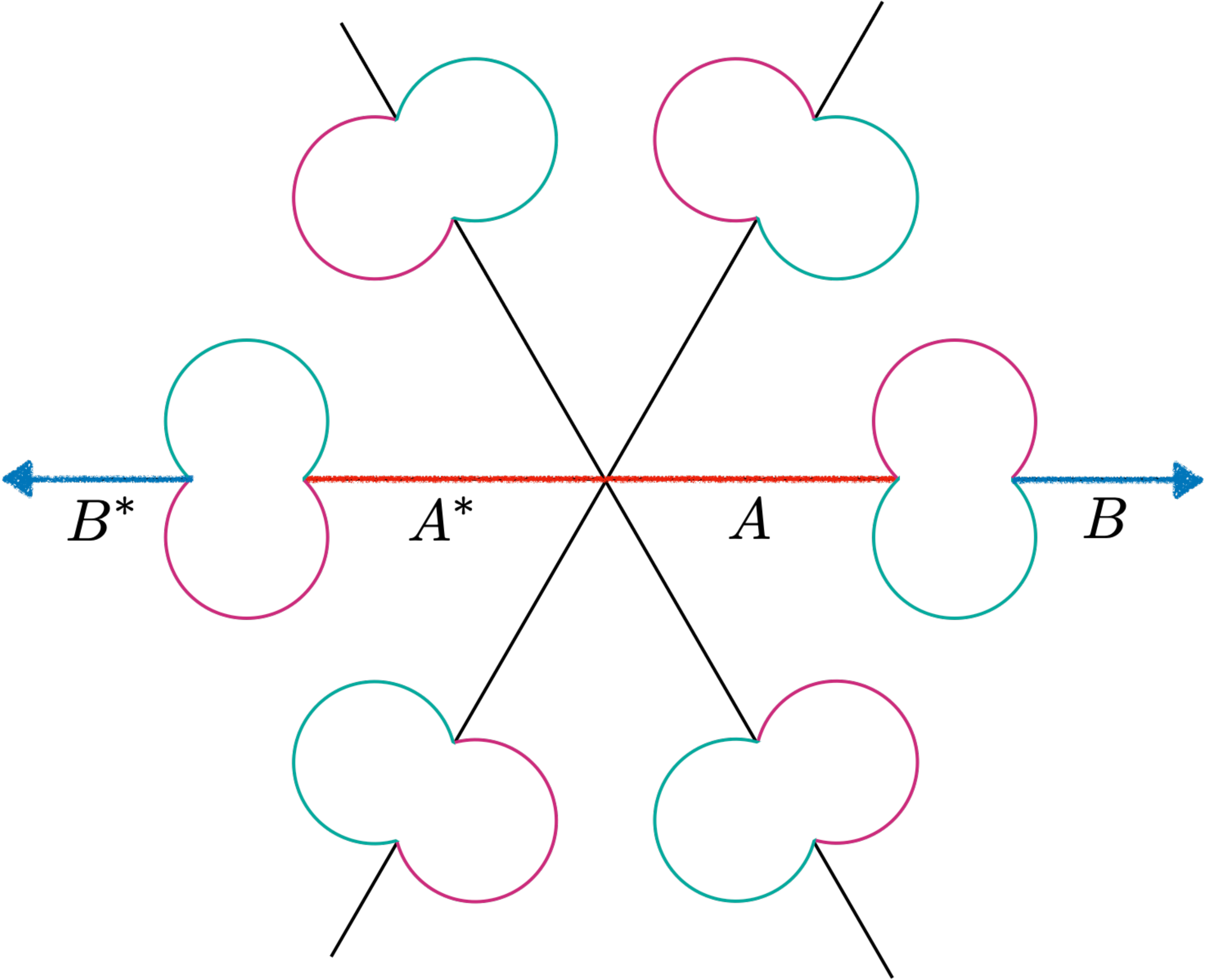}
  \caption{(left) The Schottky domain (gray) of the first replica. The intervals $A$ and $B$ are mapped to $[0,x_S]$ and $[1,\infty]$.
    The dual intervals $A^*$ and $B^*$ are mapped to $[0,e^{2\pi i/n}x_S]$ and $[e^{2\pi i/n},\infty]$.
    $x_S$ is a complicated function of the conformal cross-ratio $x=\frac{(a_1-b_1)(a_2-b_2)}{(a_1-a_2)(b_1-b_2)}$.
    Its exact form does not matter to us as we only need the fact that $x_S\to x$ when $n\to 1$.
    The two identifying circular arcs corresponds to the complementary segments $\overline{A\cup B}$ in the original replica.
    The angle between the arcs and the radial axes is $\pi/n$.
    (right) The full Schottky domain of $n=6$, produced by replicating the single Schottky domain 6 times and glue accordingly. Two adjacent arcs of different colors are identified in the same way.}
    \label{fig:schottky}
\end{figure}

In general we can write down the canonical purified density matrix $\text{Tr}_{BB^*}\ket{\rho^{n/2}_{AB}}\bra{\rho^{n/2}_{AB}}$ as a path integral on some Riemann surface.
See \cite{Dutta:2019gen} for detailed construction.
In order to apply the holographic RT formula we also need to construct a bulk solution whose conformal boundary limits to the aforementioned surface geometry.
Note that all the 3-manifolds of constant negative curvature can be expressed as a global quotient of $AdS_3$ by some discrete subgroup $\Sigma \subset PSL(2;\mathbb{C})$.
These quotient group actions, when taken limit at the conformal boundary of $AdS_3$, descends to conformal isometries on the Riemann sphere.
Therefore, if one can find a conformal mapping which maps our Riemann surface on which the path integral is defined to a single Riemann sphere $\mathbb{C}$ with quotient induced by some conformal isometric group $\Sigma'$,
the bulk solution is then easily obtained by extending action of $\Sigma'$ back into $AdS_3$.
The technology for finding such a mapping goes under the name of Schottky uniformization.
We will omit the details about how the uniformization mapping is constructed and only show the image of the map (Figure~\ref{fig:schottky}) in this appendix.
For a more complete review please refer to \cite{Faulkner:2013yia} and also \cite{Krasnov:2000zq,ZografTakhtadzhyan1988}.

\begin{figure}[h]
  \centering
  \includegraphics[width=0.6\textwidth]{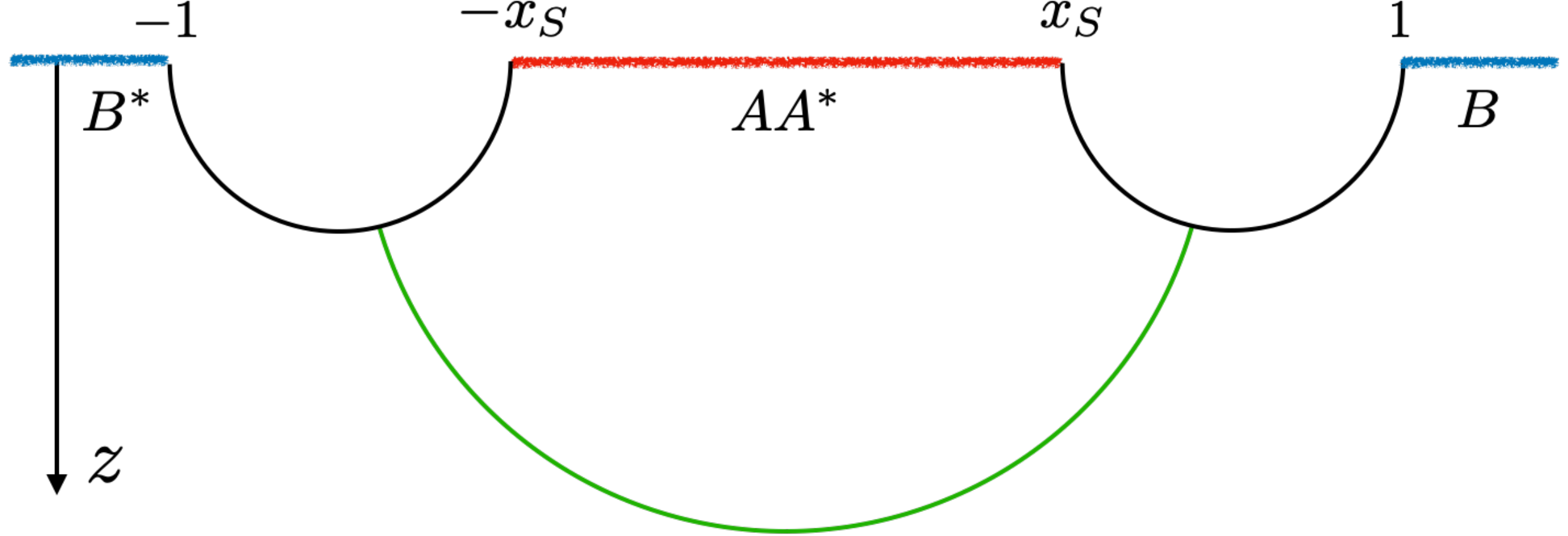}
  \caption{The bulk solution for the Schottky domain. Only a slice directly below the boundary real line is shown. The green line is the RT surface for region $AA^*$. To achieve the minimal condition the green line must meet both the circles at a perpendicular angle.}
    \label{fig:schottky2}
\end{figure}

The bulk solution for the Schottky domain is obtained by extending the circular arcs to hemispheres in the bulk.
We can then use the RT formula to find the entanglement entropy for region $AA^*$.
The holonomy condition now allows the endpoints of the minimal surface to freely move on the hemispheres, see Figure~\ref{fig:schottky2}.
The length of the RT surface can be readily found through a minimization procedure. It is
\begin{align}
  L = \ln \left| \frac{1+\sqrt{w}}{1-\sqrt{w}} \right|
\end{align}
where $w = (4x_S)/(1+x_S)^2$ is a conformal cross-ratio for the four boundary points $\{-1,-x_S,x_S,1\}$ and $x_S$ is some complicated function of the endpoints of the regions.
We will not need the detailed form of $x_S$.
What we only need is the result that $x_S\to x$ as $n\to 1$, the actual cross-ratio for the region $\{A,B\}$ on the original geometry.
We then obtain the formula of the reflected entropy for a holographic CFT
\begin{align}
  R_f = \frac{c}{3}\ln \left| \frac{1+\sqrt{x}}{1-\sqrt{x}} \right|
\end{align}

Now let's see what happens when we turn on $m$.
With nonzero and integer $m$ the picture in figure \ref{fig:schottky} is almost the same, as $m$ does not change the total replica number.
Its only effect is to shift the dual regions $A^*$ and $B^*$ in such a way that they now meet with their counterpart at a non-flat angle.
\begin{figure}[h]
  \label{fig:schottky3}
  \centering
  \includegraphics[width=0.5\textwidth]{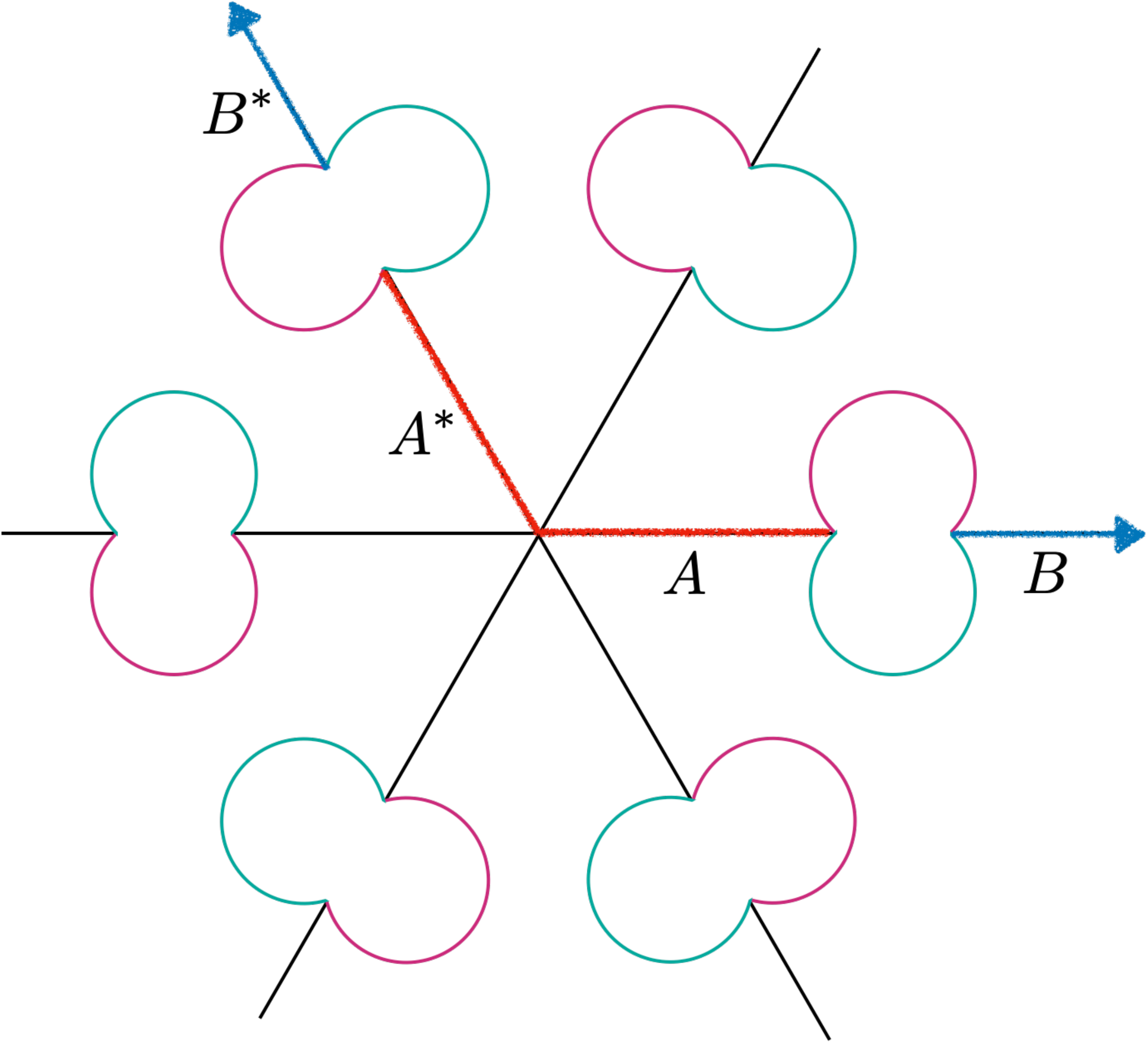}
  \caption{The Schottky domain for $n=6$ and $m=1$. Note the shift of the dual regions $A^*$ and $B^*$.}
\end{figure}
In terms of coordinates we have the following expression 
\begin{align}
  A = [0,x_S], B = [1,\infty], A^* = [0,-x_Se^{-2\pi i m/n}], B^* = [-e^{-2\pi i m/n},\infty]
\end{align}
After an analytical continuation $m\to is$ and $n\to1$ they become
\begin{align}
  A = [0,x], B = [1,\infty], A^* = [0,-xe^{2\pi s}], B^* = [-e^{2\pi s},\infty]  
\end{align}
They all lie on the real line of the boundary after the analytical continuation.
Our previous formula for the minimal surface still applies.
We only need to replace $w$ by the cross-ratio for the new points.
The result is
\begin{align}
  R_f(s) = \frac{c}{6}\ln \left| \frac{1+\sqrt{w}}{1-\sqrt{w}} \right|, \quad w = \frac{x(1+e^{2\pi s})^2}{(x+e^{2\pi s})(1+xe^{2\pi s})}
\end{align}
This expression is invariant under the reflection $s\to -s$.

\section{Next order eigenvalue correction to deflected entropy}
\label{app:B}
In this appendix we give a quick derivation of \eqref{eq:s_lambdashift}.
The deflected correlator is (assuming $z>0$ and $w<0$)
\begin{align}
  C^s(z,w) &= \frac{1}{2\pi i (z+e^{2\pi s}w)} \frac{Q_s-1}{Q_s+e^{2\pi s}} \\
  Q_s &= -\frac{z(z+1/b)}{(z+b)}\frac{e^{2\pi s}w-b}{e^{2\pi s}w (e^{2\pi s}w-1/b)}
\end{align}
For $z<0$ and $w>0$ we use $C^s(w,z) = -C^s(z,w)$. For $zw>0$ we have $C^s(w,z) = \frac{1}{2\pi i(z-w)}$.
This correlator undergoes a change $s\to -s$ after inversion $C^s(z,w) = C^{-s}(1/z,1/w)/(zw)$.
A direct consequence of this is that the contribution from two different crossed regions $(z\sim b, w\sim1/b)$ and $(z\sim 1/b,w\sim b)$ are no longer equal but evaluates to $s\to -s$.
This ensures that the eigenvalues (and hence the entropy) of $C^s$ are invariant as we invert the sign of $s$, as expected from the symmetry of modular flow.
It also greatly simplifies our work since we only need to evaluate one scaling region. 
The eigenfunctions are related to the $s=0$ case
\begin{equation}
  g^s_\nu(z) = g_{\nu^s}(e^{\pi s}z,  b\to b/\cosh(\pi s))
\end{equation}
It has a new normalization
\begin{equation}
  \int g^s_\nu(z)g^{s*}_{\nu'}(z)dz =  -4\delta_{\nu,\nu'}e^{-\pi s}\ln\left( \frac{b}{2\cosh(\pi s)}\right)
\end{equation}
Therefore the first order correction to eigenvalues are given by
\begin{equation}
  -4\ln\left( \frac{b}{2\cosh(\pi s)} \right)\delta\lambda = e^{\pi s}\int dz dw \delta C^s(z,w) g^s_\nu(z)g^{s*}_\nu(w)
\end{equation}
We have factored $e^{\pi s}$ to RHS as it restores the $s\to -s$ symmetry of the integral, as we will see later.
Again there are multiple scaling regions to consider, but as in the $s=0$ case only the crossed region $z\sim|b|, w\sim|1/b|$ and $z\sim|1/b|, w\sim|b|$ contributes.

  Consider $z>0$ and $w<0$ and $z\sim b, w\sim 1/b$. The correlator is
  \begin{equation}
    \delta C^s(z,w) = -\frac{1}{2\pi i w}\frac{(z/b+1)(e^{2\pi s}bw-1)+z/b}{(z/b+1)(e^{2\pi s}bw-1)-e^{2\pi s}z/b}
  \end{equation}
  The relavent integral is
  \begin{equation}
    \frac{(-)^m b}{2\pi} \int^\Lambda_1dx\int^\Lambda_1dy x^{-1/2-i\nu}y^{-1/2+i\nu} \frac{1}{e^{2\pi s}+xy}\left( e^{\pi s}\tanh(\pi s) + \frac{x+y}{2\cosh{\pi s}} \right)
  \end{equation}
  with $x = 1+2e^{\pi s}z\cosh(\pi s)/b$ and $y = -2\cosh(\pi s)/(e^{\pi s}wb)+1$.
  For the same sign but for $z\sim 1/b, w\sim b$ we have the integral
  \begin{align}
    \frac{(-)^mb}{2\pi} \int^\Lambda_1dx\int^\Lambda_1dy x^{-1/2-i\nu}y^{-1/2+i\nu}\frac{1}{e^{-2\pi s}+xy}\left( e^{-\pi s}\tanh(-\pi s)+\frac{x+y}{2\cosh(\pi s)} \right)
  \end{align}
  with $y = 1-2e^{\pi s}w\cosh(\pi s)/b$ and $x = 2\cosh(\pi s)/(e^{\pi s}zb)+1$.
  One can see the aformentioned symmetry of the correlator with $s\to -s$.
  
  These integrals are composed of two different pieces. The first one is absent in $s=0$:
  \begin{align}
    \begin{split}
       I_1 &\equiv \int^{\Lambda}_1 dx\int^{\Lambda'}_1 dy x^{-1/2-i\nu}y^{-1/2+i\nu} \frac{e^{\pi s}}{e^{2\pi s}+xy} + (s\to-s) \\
          &= \frac{2\tanh(\pi s)}{\nu}\sech(\pi\nu)\sin(2\pi s\nu)  + \text{divergent terms.}
    \end{split}
  \end{align}
  The second integral is similar to the $s=0$ case but with modified bounds:
  \begin{align}
    \begin{split}
       I_2 &\equiv \int^{\Lambda}_1dx\int^{\Lambda'}_1dy x^{-1/2-i\nu}y^{-1/2+i\nu}\frac{x+y}{e^{2\pi s}+xy} + (s\to -s) \\
          &= -\frac{2\sech(\pi\nu)}{1+4\nu^2}\left(\cos(2\pi s\nu)+2\nu\sin(2\pi s\nu)\tanh(\pi s)\right)  + \text{divergent terms.}
    \end{split}
  \end{align}
    Note that we have only kept the terms that is constant as we scale $\Lambda,\Lambda'\to\infty$ since we know the divergent terms must cancel across different regions from the discussion in Section~\ref{sec:corrections}.

  We conclude that we can write the correction for finite $s$ to be
  \begin{align}
        \begin{split}
    -4\ln\left( \frac{b}{2\cosh(\pi s)} \right)\delta\lambda^s &= (-)^mb(I_1+I_2) \\
    &=(-)^mb\left( \frac{\sech(\pi \nu)}{1+4\nu^2} \right)\left( -2\cos(2\pi s \nu) + \frac{\tanh(\pi s)\sin(2\pi s\nu)}{\nu} \right)
    \end{split}      
  \end{align}

\bibliography{biblio}

\end{document}